\documentclass[useAMS,usenatbib]{mn2e}

\usepackage{times}
\usepackage{amsmath}    % need for subequations
\usepackage{graphicx}   % need for figures
\usepackage{verbatim}   % useful for program listings
\usepackage{color}      % use if color is used in text
\usepackage{subfigure}  % use for side-by-side figures
\usepackage{amssymb}
\usepackage{cite}
\usepackage{setspace}
\newcommand{\vect}{\mathbf}

\newcommand{\threej}[6]{\left(\begin{array}{ccc} #1 & #2 & #3 \\
                                                 #4 & #5 & #6
                        \end{array}\right)}
\newcommand{\be}{\begin{equation}}
\newcommand{\ee}{\end{equation}}
\newcommand{\ba}{\begin{eqnarray}}
\newcommand{\ea}{\end{eqnarray}}
\newcommand{\bnum}{\begin{enumerate}}
\newcommand{\enum}{\end{enumerate}}

\newcommand{\bom}{\bomega}

\title[Removing beam asymmetry bias in CMB experiments]{Removing beam
  asymmetry bias in precision CMB temperature and polarisation experiments}
\author[Christopher G. R. Wallis et al.]{\parbox[t]{\textwidth}{Christopher G. R. Wallis$^1$\thanks{E-mail:
cwallis@jb.man.ac.uk}%; mbrown@jb.man.ac.uk (MLB); richard.battye@manchester.ac.uk (RAB); 
, Michael L. Brown$^1$, Richard A. Battye$^1$, \\ Giampaolo Pisano$^1$ and Luca Lamagna$^2$}\vspace*{6pt}\\
$^1$Jodrell Bank Centre for Astrophysics, School of Physics and Astronomy, The University of Manchester, Manchester M13 9PL\\
$^2$Dipartimento di Fisica, Sapienza Universit\`a di Roma, Piazzale Aldo Moro 5, 00185 Roma, Italy}
\begin{document}

\date{Accepted 2014 XXXXX XX. Received 2014 XXXXX XX; in original form 2014 XXXXX XX}

\pagerange{\pageref{firstpage}--\pageref{lastpage}} \pubyear{2014}

\maketitle

\label{firstpage}

\begin{abstract}
Asymmetric beams can create significant bias in estimates of the power
spectra from CMB experiments. With the temperature power spectrum many
orders of magnitude stronger than the B-mode power spectrum any
systematic error that couples the two must be carefully controlled
and/or removed. Here, we derive unbiased estimators for the CMB
temperature and polarisation power spectra taking into account general
beams and general scan strategies. A simple consequence of asymmetric
beams is that, even with an ideal scan strategy where every sky pixel is
seen at every orientation, there will be residual coupling from
temperature power to B-mode power if the orientation of the beam
asymmetry is not aligned with the orientation of the
co-polarisation. We test our correction algorithm on simulations of
two temperature-only experiments and demonstrate that it is
unbiased. The simulated experiments use realistic scan strategies,
noise levels and highly asymmetric beams. We also develop a map-making
algorithm that is capable of removing beam asymmetry bias at the map
level. We demonstrate its implementation using simulations and show
that it is capable of accurately correcting both temperature and
polarisation maps for all of the effects of beam asymmetry including
the effects of temperature to polarisation leakage.
\end{abstract}


\section{Introduction}\label{intro}

The cosmic microwave background (CMB) has proved to be an incredibly
useful tool for testing cosmological models. The CMB two-point
correlation functions, or alternatively in Fourier space the power
spectra, of the temperature and polarisation fluctuations are of
particular interest. There has been a wealth of experiments which have
been successful at characterising the CMB power spectra and these
observations are consistent with the standard $\Lambda$CDM
cosmological model (see e.g.~\citealt{2013arXiv1303.5076P}). The
temperature power spectrum is now extremely well characterised
\citep{2013arXiv1303.5062P} and ongoing and future experiments will
focus on characterising the polarization more accurately. 
 
The polarisation of the CMB, being a spin-2 field, can be decomposed
into curl-free ($E$-mode) and gradient-free ($B$-mode)
components. $E$-modes have been successfully detected and
characterised, and have helped to constrain
cosmology~\citep{2002Natur.420..772K, 2004Sci...306..836R,
  2006ApJ...647..813M, 2009ApJ...705..978B, 2010ApJ...711.1123C,
  2012arXiv1207.5562Q}. The fainter $B$-mode signal presents a much
greater challenge for experimentalists, though the first tentative
detections of $B$-mode polarization on small scales are now being made
through cross-correlations~\citep{2013PhRvL.111n1301H,
  2013arXiv1312.6646P}.

The $E$-mode power spectrum is approximately two orders of magnitude
fainter than the temperature power spectrum while the $B$-mode power
spectrum is expected to be at least 2 orders of magnitude fainter still
\citep{2013IAUS..288...42C}. This means that in any experiment aimed
at detecting $B$-modes, systematic effects that could potentially
couple the temperature or $E$-mode signal to the $B$-mode power
spectrum must be strictly controlled. One source of potential error is
an asymmetric optical response function (i.e.~the experimental
beam). During the data analysis for CMB experiments, one often assumes
that the beam is axisymmetric so that its effect is to simply (and
isotropically) smooth the sky. This assumption can result in a bias in
subsequent power spectrum estimates if, as is often the case in
reality, the beam is not perfectly axisymmetric.

At present this has not been crucial in extracting CMB power spectra,
but with increased sensitivity of instruments, this bias will be
important for future experiments. Currently there are three approaches in the
literature that attempt to deal with this effect. The first is 
simply to quantify the systematic error on the cosmological parameters
caused by the asymmetry and be satisfied that they are below the
statistical uncertainty. This can be done by simulating an experiment's
full beam response~\citep{2011ApJS..193....5M} and propagating the
error~\citep{2013arXiv1303.5068P}. This approach is effective as the error
in the maps can be estimated. {\color{black} In their analysis the {\it Planck} 
team accounted for the effects of beam asymmetry by finding the most suitable axisymmetric 
effective beam transfer function to deconvolve the recovered power spectrum.
Any further asymmetry bias was shown to have little to no effect on the science
\citep{2013arXiv1303.5062P}.} The
second method investigated attempts a full deconvolution of the
Time Ordered Data (TOD) from a CMB experiment to remove the effect
completely~\citep{2001PhRvD..63l3002W, 2000PhRvD..62l3002C, 2012A&A...548A.110K}. The results of this method are
encouraging but it is not able to deal with certain unavoidable real-world
complications. In particular when noise is added to the TOD the
deconvolution no longer works for high multipoles. In addition, the
deconvolution only works if the experiment observes the entire
sky. This is not a significant problem for satellite-based experiments
but for ground- or balloon-based experiments this will obviously not
be the case. 

A third approach is to calculate, and subsequently correct for, the
asymmetry bias on the measured pseudo-$C_{\ell}$ in an
experiment~\citep{2013arXiv1309.4784R,2006NewAR..50.1030S}. As
presently formulated this is unable to deal with a cut sky without
apodising the azimuthal dependence of the mask. This in turn results in a 
reduction of the cosmological information content of the TOD which is
something that we would like to avoid. In addition, these authors
assume that each sky pixel is seen in a single orientation only,
whereas, in general, experiments will observe each sky pixel in a
number of different orientations. 
%Both these short-comings mean that, at present, this method could not
%be applied to real observations.

In this paper, we present two methods to remove the effect of
asymmetry bias. The first is an algorithm to recover the CMB power
spectra using the pseudo-$C_{\ell}$ approach. We make no assumptions
about the beam or the scan strategy in developing unbiased estimators
for the underlying CMB temperature and polarisation power spectra. We
also show that noise can be easily accommodated. The pseudo-$C_{\ell}$
estimator that we propose is based on a calculation similar to one
presented in \citet{2010PhRvD..81j3003H}. Here, we extend the analysis
to polarisation and demonstrate its implementation on simulated
temperature-only experiments. Since the estimator works directly on
the time ordered data (TOD), it is sub-optimal for polarization
experiments that do not directly measure both $Q$ and $U$ Stokes
parameters in the timeline (e.g.~via detector differencing). For such
experiments, the estimator will perform the decomposition into $Q$ and
$U$ at the level of the power spectrum which will contribute to the
statistical error. However, this estimator is well suited to an
experiment such as {\it Planck} which has both instrument-$Q$ and
instrument-$U$ detector pairs on its focal plane.

In addition to the pseudo-$C_\ell$ approach, we present a new
map-making algorithm that is capable of making temperature and
polarisation maps cleaned of asymmetry bias. The map-making algorithm
produces maps containing the sky signal smoothed with just the
axisymmetric components of the beam and noise. Note that this
map-making scheme requires a suitable scan strategy --- in general,
the more complex the beam asymmetry is, the more redundancy (in terms
of polarization angle covereage) is required in the scan strategy. For the
asymmetric beams that we have investigated in this paper, the scan
strategy requirements are fully met by scanning modes proposed for
future CMB satellite experiments such as those described in
\citet{2009arXiv0906.1188B}.

Removing the asymmetry at the map level has two main
benefits. Firstly, in contrast to the case of the pseudo-$C_\ell$
approach, the polarisation power spectrum estimator error bars are not
affected by the sample variance of the temperature power
spectrum. Secondly, the resulting bias-free temperature and
polarisation maps can also be used for science other than power spectrum
estimation. Foregrounds can be removed after the map-making has been
performed meaning that current component separation techniques can be
applied.

The paper is organized as follows. We begin in
Section~\ref{sec:setupprob} where we present some basic definitions
and develop the mathematical formalism on which our algorithms are
based. In Section~\ref{sec:pseudo-cl-technique}, we present the
pseudo-$C_\ell$ based approach to correcting for beam
asymmetry. Section~\ref{sec:TtoPleak} discusses the potential
impact of beam asymmetries on CMB polarization experiments, if they
are left uncorrected. In Section~\ref{sec:map_making}, we present 
a technique to correct for the effects of beam asymmetry in the map
domain. Section~\ref{sec:reducesum} discusses some details of the
decomposition of the beam which is required for both of our
approaches. We demonstrate our techniques on simulations in
Sections~\ref{sec:sim_test} \& \ref{sec:sim_test_map} and our conclusions
are presented in Section~\ref{sec:discuss}.

%\section{Setting up the problem}\label{sec:setupprob}
\section{Basic definitions and preliminaries}\label{sec:setupprob}
%\subsection{Preliminaries}
Our objective is to construct estimators for the temperature and
polarisation fluctuation power spectra given a TOD. We assume that any 
non-astrophysical signals in the TOD have been flagged and, for the
pseudo-$C_\ell$ approach, that foregrounds have been removed and/or
masked.  We consider an asymmetric beam and a general scan
strategy. We begin by defining some relevant quantities.

The CMB temperature and polarisation fluctuations, $\Delta
T(\theta,\phi), Q(\theta,\phi)$ and $U(\theta,\phi) $, can be
decomposed into spin-weighted spherical harmonics
\ba a_{0\ell m} &=& \int \: d\Omega\, _0Y^*_{\ell m}(\Omega)\Delta
T(\Omega) \:\:\:\text{and} \\ a_{\pm2\ell m} &=& \int \: d\Omega\,
_{\pm2}Y^*_{\ell m}(\Omega)[Q(\Omega) \mp i U(\Omega)] , 
\ea 
where $_{s}Y_{\ell m}$ are the spin weighted spherical harmonics. The
temperature, $E$-mode and $B$-mode harmonic coefficients  are
related to these by
\ba
a_{\ell m}^T &=& a_{0 \ell m},\\
a_{\ell m}^E &=& -\frac{1}{2}\left(a_{2\ell m}+a_{-2\ell m}\right), \label{eq:almpm2toE}\\
a_{\ell m}^B &=& -\frac{1}{2i}\left(a_{2\ell m}-a_{-2\ell m}\right). \label{eq:almpm2toB}
\ea
We are interested in obtaining unbiased estimates for the power and
cross spectra of the CMB defined as
\ba
C^{XY}_{\ell} = \frac{1}{2\ell+1}\sum_{m=-\ell}^{\ell}a^{X}_{\ell m}a^{Y*}_{\ell m}, \label{eq:pcspectra}
\ea
where $X,Y {=} \{ T,E,B \} $. 

The response of a telescope to the Stokes parameters on the sky
$(T,Q,U)$ can be described by some response (beam) functions
$(\tilde{T},\tilde{Q},\tilde{U})$. The total detected power is
\ba
W \propto \int (T\tilde{T}+Q\tilde{Q} + U\tilde{U})d\Omega.
\ea
For details of how polarised detectors respond to the CMB see
\citet{2000PhRvD..62l3002C}. One element of the TOD
is then this power $W$ integrated over the time interval between two
samples. We define the spin weighted spherical harmonic transforms of
the beam to be
\ba
b_{0\ell k} &=& \int \: d\Omega\, _0 Y^*_{\ell
  k}(\Omega)\tilde{T}(\Omega) \label{eq:b_decomp_s0} \:\:\:\text{and} \\
b_{\pm2\ell k} &=& \int \: d\Omega\, _{\pm2}Y^*_{\ell
  k}(\Omega)[\tilde{Q}(\Omega) \mp i \tilde{U}(\Omega)] \label{eq:b_decomp_s2},
\ea
when the beam is pointing in the $z$-direction in a fiducial
orientation such that the co-polarisation is aligned with the
$y$-direction. Note that this formalism can describe all aspects of a
detector's response function. For example, we can include both the
asymmetry of the beam and any cross-polarisation response. This will
allow us to remove any bias that these beam imperfections would impart
on the estimated power spectra.

We truncate our expansion of both the beam and the sky at some maximum
multipole, $\ell_{\rm{max}}$. We also cap the expansion of the beam in
$k$ at some maximum value, $k_{\rm{max}}$. This is a reasonable
approximation to make as beam response functions are typically close
to axisymmetric. In Section~\ref{sec:reducesum} we will examine this
assumption for some specific cases. The beam is then rotated around
the sky in a scan to measure the CMB. We describe this rotation using
Euler angles {\color{black} $\bom {=}[\phi,\theta,\psi]$.} This is really three active
rotations. They are active as the beam moves with respect to the
coordinate system. The following series of steps describe how to
rotate the beam from the fiducial orientation to the orientation
described by $\bom$, all rotations being perfomed anticlockwise when
looking down the axis by which they are defined.

\bnum
\item The beam is rotated around the $z$ axis by $\psi$.
\item The beam is rotated by $\theta$ around the $y$ axis.
\item The beam is rotated around the $z$ axis again by $\phi$.
\enum

The Wigner D-matrix, $D^{\ell}_{mk}(\bomega)$, performs the required
rotations on the spherical harmonic decomposition of a
function. Therefore we can write one element of the TOD ($t_j$) as
\ba
t_j &=& \sum_{s\ell mk} D^{\ell*}_{mk}(\bomega_j)b^*_{s\ell k}a_{s\ell m}.\label{eq:td}
\ea
For simplicity, in this paper, we only write the index which is being
summed over and not the ranges. For the rest of this paper one should
assume $\ell$ ranges from 0 to $\ell_{\rm{max}}$, $m$ from $-\ell$ to
$\ell$ and $s{=}0,\pm2$. The index $k$ ranges from $-k_{\rm{max}}$ to
$k_{\rm{max}}$, unless $\ell < k_{\rm{max}}$, in which case the range is the
same as for $m$.

Before we can go further we must define some more mathematical
constructs. The first is the ``hit cube'', $W(\bom) {\equiv} \sum_j
\delta(\bom-\bom_j)$, which describes which sky positions the
experiment has observed, and in which orientations. $W(\bom)$ is
defined on the space running from 0 to $\pi$ in $\theta$ and from 0 to
$2\pi$ in both $\phi$ and $\psi$. The infinitesimal volume element of
this space $d^3\bom{=}\sin\theta d\theta \, d\phi \, d\psi$.

The convolution of the sky with the beam $t(\bom)$ is a continuous
function which we only have limited knowledge of. The knowledge we
have is dictated by the scan strategy, or in this formalism the hit
cube. Formally we have
\ba
t(\bom) &=& \sum_{s \ell m k} D^{\ell *}_{ m k}(\bom)b^*_{s \ell k}a_{s \ell m},\label{eq:t_conv}\\
\tilde{t}(\bom) &=& W(\bom)t(\bom).
\ea
$\tilde{t}(\bom)$ is a function which contains all of the
astrophysical information present in the experiment. It is simply a
rewriting of the TOD, where each element is decribed by a
$\delta$-function at the relevant orientation $\bom$. 

The Wigner D-matrices provide a complete orthogonal basis for this
space, so we use them to decompose both the TOD and the hit cube as
\ba
T_{mk}^{\ell*} &=& \int d^3\bom D^{\ell}_{mk}(\bom)\tilde{t}(\bom)n(\bom) \:\:\:\text{and} \label{eq:Tlmk} \\
w^\ell_{mk} &=& \frac{2\ell +1}{8\pi^2}\int d^3\bom D^{\ell*}_{mk}(\bom)W(\bom)n(\bom). \label{eq:wlmk}
\ea
Here, we have introduced the weighting function $n(\bom)$, which
ranges from 0 to 1. We use this function to down-weight noisy pixels,
apply a foreground mask and/or apodise the hit cube so that is can be
described well within our expansion. \footnote{{\color{black} Specifically
we consider a weighting function that openly depends on $\theta$ and 
$\phi$. Our choice of normalised weighting function for a given sky pixel
is $n(p) {=} m(p)/ N _{\rm hits}(p)$,
where $m(p)$ is the mask applied to pixel $p$ and $N_{\rm hits}(p)$ is the 
number of hits that pixel $p$ receives in the scan strategy.
Weighting a TOD element with a factor of $1/N_{\rm hits}(p)$ is equivalent
to the weight the element receives when making a binned map and should
not be considered to be down weighting high signal to noise regions of the map.}}
 We use the prefactor in
equation~\eqref{eq:wlmk} to correctly normalise the coefficients so
that we can write
\ba
W(\bom)n(\bom) = \sum_{\ell mk} w^\ell_{mk}D^\ell_{mk}(\bom).
\ea
For this reconstruction of the hit cube to be exact, one would require
$k$ to range from $-\ell$ to $\ell$, not as we have here from $-k_{\rm
  max}$ to $k_{\rm max}$. This is not a problem for our proposed
techniques since, as we shall see later, we only need to capture the
features in the $\psi$ direction with Fourier modes up to some
$k_{\rm{max}}$, the value of which is determined by the complexity of
the beam asymmetry. This is analogous to the axisymmetric case, where
to recover the temperature fluctuation power spectrum the analysis
uses the hit map, which is just the $k{=}0$ Fourier mode of the hit
cube. So while we cannot fully recover the complete hit cube we do
recover its important features.

\section{Beam asymmetry correction within the pseudo-$C_\ell$ framework}
\label{sec:pseudo-cl-technique}
In this section, we develop a technique to correct for the effects of
beam asymmetry within the framework of pseudo-$C_\ell$ power spectrum
estimators~\citep{2002ApJ...567....2H,
  2005MNRAS.360.1262B}. As
described earlier, this approach to correcting for beam asymmetries
is well suited to experiments that can measure the $Q$ and $U$
Stokes parameters simultaneously in the timestream (e.g.~differencing
experiments). In Appendix \ref{sec:rel_to_stand} we show that for such an experiment the 
pseudo-$C_\ell$ presented here is similar to that of the standard 
pseudo-$C_\ell$ approach \citep{2005MNRAS.360.1262B}.

\subsection{Definition of the pseudo-$C_\ell$ spectra}
We wish to define a two-point statistic that contains the relevant information
from the TOD and which can also be related to the power spectra defined in
equation~\eqref{eq:pcspectra} via a coupling matrix. As the quantity
$T^\ell_{mk}$ contains all of the information present in the TOD, it
must therefore contain all of the information required to recover the CMB
spectra. We define the pseudo-$C_\ell$ spectra as

\ba
\tilde{C}_{\ell}^{kk'} &\equiv& \sum_m T^{\ell*}_{mk} T^{\ell}_{mk'}, \label{eq:pscl}
\ea
which can be computed directly from the TOD. {\color{black} The appropriateness of
this choice becomes clear when we write the Wigner D-matrices in terms
of the spin weighted spherical harmonics, from equation (3.10) in 
\citet{1967JMP.....8.2155G}:
\ba
D^{\ell}_{m k}(\phi,\theta,\psi) = \sqrt{\frac{4\pi}{2\ell+1}}    e^{ik\psi}~_{-k}Y_{\ell m}(\theta, \phi). \label{eq:wig_spin}
\ea
} We see that $T^{\ell}_{m0}$ will be similar to the $a_{0\ell m}$
coefficients of a binned map made from the TOD while $T^\ell_{m,\pm2}$
will be similar to the $a_{\pm2\ell m}$ coefficients. This is due to the
fact that the Wigner D-matrices are decomposing the 3D space, $\bom$,
with basis functions over the $\theta$ and $\phi$ dimensions which are
the spin weighted spherical harmonics.

\subsection{Calculating the coupling operator}\label{sec:calcoup}
Here we aim to find an analytic expression for the
$\tilde{C}_\ell^{kk'}$ defined in equation \eqref{eq:pscl} in terms of
the true sky spectra and a coupling matrix that depends only on the
scan strategy and the beam. We begin by re-writing the decomposition
of the TOD using equation (\ref{eq:td}) and the definition of the
window function. We do this to replace the sum over $j$
with an integral,
{\color{black}\ba
T^{\ell_1*}_{m_1 k_1} &=& \sum_j  D^{\ell_1}_{m_1 k_1} (\bomega_j)t_jn(\bom_j)\nonumber \\
&=&\!\!\!\!\!\!\!\sum_{j s_2 \ell_2 m_2 k_2} \!\!\!\!\!\!\!D^{\ell_1}_{m_1 k_1} (\bomega_j)D^{\ell_2*}_{m_2 k_2}(\bomega_j)b^*_{s_2 \ell_2 k_2}a_{s_2 \ell_2 m_2}n(\bom_j). \nonumber 
\ea
} Using the definition of the hit cube $W(\bom){=}\sum_j
\delta(\bom-\bom_j)$ we deduce that
%&=&\!\!\!\!\!\!\!\sum_{s_2 \ell_2 m_2 k_2} \!\!\!\!\!b^*_{s_2 \ell_2 k_2}a_{s_2 \ell_2 m_2} \nonumber \\
%&\:&\:\:\:\:\:\:\:\:\times\int d^3\bom D^{\ell_1}_{m_1 k_1} (\bomega)D^{\ell_2*}_{m_2 k_2}(\bomega)W(\bomega)n(\bom)\nonumber \\
\ba
T^{\ell_1*}_{m_1 k_1} &=&\!\!\!\!\!\!\!\sum_{s_2 \ell_2 m_2 k_2} \!\!\!\!\!b^*_{s_2 \ell_2 k_2}a_{s_2 \ell_2 m_2} K_{m_1 k_1 m_2 k_2}^{\ell_1 \ell_2},\label{eq:TfromK}
\ea
%
%\ba
%T^{\ell_1*}_{m_1 k_1} &=& \sum_j  D^{\ell_1}_{m_1 k_1} (\bomega_j)t_j\nonumber \\
%&=&\sum_{j \ell_2 m_2 k_2} \!\!\!\!\!D^{\ell_1}_{m_1 k_1} (\bomega_j)D^{\ell_2*}_{m_2 k_2}(\bomega_j)b^*_{\ell_2 k_2}a_{ell_2 m_2} \nonumber \\
%&=&\sum_{\ell_2 m_2 k_2} \!\!\!b^*_{\ell_2 k_2}a_{\ell_2 m_2} \nonumber \\ 
%\:&\:\:\times \int d^3\bom D^{\ell_1}_{m_1 k_1} (\bomega)D^{\ell_2*}_{m_2 k_2}(\bomega)W(\bomega)\nonumber \\
%&=&\sum_{\ell_2 m_2 k_2} \!\!\!b^*_{\ell_2 k_2}a_{\ell_2 m_2} K_{m_1 k_1 m_2 k_2}^{\ell_1 \ell_2},
%\ea
%
where in the last line we have introduced the coupling kernel,
\ba
K_{m_1 k_1 m_2 k_2}^{\ell_1 \ell_2} &\equiv \int d^3 \bom D^{\ell_1}_{m_1 k_1}(\bomega) D^{\ell_2 *}_{m_2 k_2}(\bomega) W(\bomega)n(\bom).
\ea
We are now in a position to calculate the coupling operator. We start
from the definition of $\tilde{C}_\ell^{k_1 k_1'}$,
\ba
\tilde{C}_{\ell_1}^{k_1 k_1'} &\equiv& \sum_{m_1}T^{\ell_1*}_{m_1 k_1}T^{\ell_1}_{m_1 k_1'}\nonumber \\
&=& \!\!\!\!\!\!\sum_{\substack{m_1{}\\ s_2 \ell_2 m_2 k_2{}\\s_3 \ell_3 m_3 k_3}}\!\!\!\!\!\!b^*_{s_2 \ell_2 k_2}a_{s_2 \ell_2 m_2} K_{m_1 k_1 m_2 k_2}^{\ell_1 \ell_2}\nonumber\\
&& \:\:\:\:\:\:\:\:\:\:\:\:\:\times \:\:b_{s_3 \ell_3 k_3}a^{*}_{s_3 \ell_3 m_3} K_{m_1 k_1' m_3 k_3}^{\ell_1 \ell_3*}.
\label{eq:cl2pscl_prelim}
\ea
If we now assume that the CMB temperature and polarization fluctuations are
Gaussian distributed with isotropic variance, then we can write $\langle a_{s
  \ell m}a^*_{s' \ell'm'}\rangle =  C_\ell^{ss'} \delta_{\ell\ell'}\delta_{mm'}$ 
where we have defined
\ba
C^{s s'}_{\ell} = \frac{1}{2\ell+1}\sum_m a_{s\ell m} a^*_{s' \ell
  m}. \label{eq:spin_spec}
\ea
{\color{black} Using this result, equation
~\eqref{eq:cl2pscl_prelim} simplifies to
\ba
\langle \tilde{C}_{\ell_1}^{k_1 k_1'}\rangle &=& \!\!\!\!\!\!\!\sum_{\substack{m_1 {}\\ s_2 \ell_2 m_2 k_2{}\\ s_3 k_3}}\!\!\!\!\!\!b^*_{s_2 \ell_2 k_2} K_{m_1 k'_1 m_2 k_2}^{\ell_1 \ell_2}b_{s_3 \ell_2 k_3} K_{m_1 k'_1 m_2 k_3}^{\ell_1 \ell_2*} C^{s_2 s_3}_{\ell_2} \nonumber \\
&=& \!\!\!\!\sum_{\substack{s_2 \ell_2 k_2{}\\ s_3 k_3}} \!\!\!b_{s_2 \ell_2 k_2}^* b_{s_3 \ell_2 k_3} M^{\ell_1 \ell_2}_{k_1 k_1' k_2 k_3} C_{\ell_2}^{s_2 s_3}, \label{eq:cl2pscl}
\ea
where, in the second step, we have used the product of two coupling kernels, $M_{k_1
  k'_1 k_2 k_3}^{\ell_1 \ell_2}$, derived in
Appendix~\ref{sec:productkern}. We can now identify the coupling
operator that describes the contribution of each sky spectrum
$C_{\ell}^{s_1 s_2}$ to each of the $\tilde{C}_\ell^{k k'}$, i.e. we
can write 
\ba
\langle \tilde{C}_{\ell_1}^{k_1 k_1'}\rangle &=& \sum_{\ell_2 s_2 s_3} O_{\ell_1 \ell_2}^{k_1 k_1' s_1 s_2} C_{\ell_2}^{s_2 s_3}, \label{eq:red_cl2pscl}
\ea}
where,
\ba
O_{\ell_1 \ell_2}^{k_1 k_1' s_1 s_2} &=& \sum_{k_2 k_3} b_{s_2 \ell_2 k_2}^* b_{s_3 \ell_2 k_3} M^{\ell_1 \ell_2}_{k_1 k_1' k_2 k_3}.\label{eq:oper}
\ea
Certain symmetries can be used to reduce the number of $M$ matrices
required to evaluate equation~\eqref{eq:oper}. These symmetries, which
are derived in Appendix~\ref{sec:symM}, are
\ba
M^{\ell_1 \ell_2}_{k_1 k_1' k_2 k_3} &=& M^{\ell_1 \ell_2*}_{k'_1 k_1 k_3 k_2},\\
M^{\ell_1 \ell_2}_{k_1 k_1' k_2 k_3} &=& (-1)^{k_2 - k_1 +k_3 - k'_1}M^{\ell_1 \ell_2*}_{-k_1 -k_1' -k_2 -k_3}.
\ea

\subsection{Recovering the true CMB spectra} \label{sec:est_cmb}
We are now in a position to obtain unbiased estimators for the true
CMB power spectra $C_{\ell}^{XY}$ defined in equation
\eqref{eq:pcspectra}. We start by defining a large vector made up of
the pseudo-$C_\ell$ spectra, $\tilde{C}_\ell^{kk'}$, and another
comprised of the true full-sky spin spectra, $C_\ell^{ss'}$ defined in
equation~\eqref{eq:spin_spec}:
\ba
\vect{C}_{i} &=&
(C_\ell^{00},C_\ell^{02},C_\ell^{0-2},C_\ell^{22},C_\ell^{2-2},C_\ell^{-2-2})^{T}
\label{eq:big_C_defn1}\\
\vect{\tilde{C}}_{i} &=& (\tilde{C}_\ell^{00},\tilde{C}_\ell^{02},\tilde{C}_\ell^{0-2},\tilde{C}_\ell^{22},\tilde{C}_\ell^{2-2},\tilde{C}_\ell^{-2-2})^{T}.
\label{eq:big_C_defn2}
\ea
Each of these are vectors of length $6(\ell_{\rm max} +1)$. Using 
these definitions we can write our overall coupling matrix equation as
\ba
\vect{\tilde{C}}_{i_1} &=& \sum_{i_2} \vect{O}_{i_1 i_2} \vect{C}_{i_2}. \label{eq:polar_op}
\ea
We write the overall coupling operator $\vect{O}_{i_1 i_2}$ explicitly
in terms of the individual $O$-matrices of equation~\eqref{eq:oper} in
Appendix~\ref{app:coup_op}. Once this operator has been calculated,
equation~\eqref{eq:polar_op} can be inverted to recover the true spin
spectra $C^{ss'}_\ell$, properly deconvolved for both the mask and the
asymmetric beam. A further simple transformation, which is explicitly
written down in Appendix~\ref{app:spin2XY}, yields the final estimates
of the six possible CMB power spectra,  $C^{XY}_\ell$. Note that, as
with normal pseudo-$C_\ell$ estimators, in the presence of a severe
sky cut, the matrix, $O_{i_1 i_2}$ will be singular and must therefore
be binned before it can be inverted. This is standard practice with
pseudo-$C_\ell$ power spectrum estimators~\citep{2002ApJ...567....2H,
  2005MNRAS.360.1262B}.

%With this coupling operator calculated, we can now examine the effect
%that the asymmetry has on the measured E- and B-mode power spectra. As
%$T^{\ell}_{m\pm2}$ are related to the $a_{\pm2 \ell m}$ from a map
%made using the TOD, the $\tilde{C}_{\ell}^{EE}$,
%$\tilde{C}_{\ell}^{BB}$ etc. defined in \citet{2005MNRAS.360.1262B}
%are related to the $\tilde{C}_{\ell}^{k k'}$. Therefore we can examine
%the effect the asymmetry of the beam has on the measured E- and B-mode
%power spectra if it was not accounted for by defining,
%\ba
%T_{mE}^{\ell} &\equiv& -\frac{1}{2}\left(T_{m2}^{\ell}+T_{m-2}^{\ell} \right) \\
%T_{mB}^{\ell} &\equiv& -\frac{1}{2i}\left(T^{\ell}_{m2}-T^{\ell}_{m-2}\right)\:\: \text{and}\\
%\implies \tilde{C}_{\ell}^{EE} &=&
%\frac{1}{4}\left(\tilde{C}_{\ell}^{22} + \tilde{C}_{\ell}^{-2-2} 
%+ \tilde{C}_{\ell}^{-22} + \tilde{C}_{\ell}^{2-2}\right) \\
%\implies \tilde{C}_{\ell}^{BB} &=&
%-\frac{1}{4}\left(\tilde{C}_{\ell}^{22} + 
%\tilde{C}_{\ell}^{-2-2} - \tilde{C}_{\ell}^{-22} - \tilde{C}_{\ell}^{2-2}\right).
%\ea
%Therefore we examine the effect of the asymmetry and cross polar
%response of the beam by calculating the appropriate coupling
%operators.

\subsection{Including noise}
\label{sec:pseudo_cl_noise}
Any useful algorithm for removing the effects of beam asymmetry must
also be able to deal with instrumental noise. Since our algorithm
works within the framework of the standard pseudo-$C_\ell$ 
technique, we can use exactly the same approach to remove the noise
bias as is adopted in the standard
analysis~\citep{2002ApJ...567....2H}. A TOD element including noise
can be written as 
\ba
t_j &=& \sum_{slmk} D^{\ell*}_{mk}(\bomega_j)b^*_{slk}a_{slm} + n_j.
\ea
If we assume that the noise is not correlated with the pointing direction of
the telescope then
\ba
(\tilde{C}_{\ell}^{kk'})^{\rm SN}= (\tilde{C}_{\ell}^{kk'})^{\rm S} + N_\ell^{kk'}, \label{eq:noise_power}
\ea
where the ${\rm SN}$ and ${\rm S}$ superscripts denote the signal-plus-noise and
the signal-only pseudo-$C_\ell$ spectra respectively. An unbiased
estimate of the noise power spectra, $N_\ell^{kk'}$, can be obtained
by performing a set of simulations containing only instrument noise and calculating
$\tilde{C}_{\ell}^{kk'}$ for each as before using equation~\eqref{eq:pscl}. 
The noise bias is then estimated as the average over the set of noise
realisations, $N_\ell^{kk'} = \langle
C_{\ell}^{kk'} \rangle$. The final estimator for the full-sky,
noise-debiased and asymmetry-cleaned spin power spectra
can then be written as 
\ba
\vect{C}_{i_1} = \sum_{i_2} \vect{O}^{-1}_{i_1 i_2}
(\vect{\tilde{C}}_{i_2} - \langle \vect{N}_{i_2} \rangle), 
\ea
where $\langle \vect{N}_{i_2} \rangle$ is a vector of length
$6(\ell_{\rm max} + 1)$ comprised of all of the individual noise bias
spectra, constructed in an analgous fashion to
equations~\eqref{eq:big_C_defn1} and \eqref{eq:big_C_defn2}. As
before, the six CMB power spectra are then recovered trivially using
the relation in Appendix~\ref{app:spin2XY}.

\section{Impact of beam asymmetries on CMB polarization experiments}
\label{sec:TtoPleak}
With the anaysis of the previous two sections in place, we can now
examine the effect that beam asymmetries will have on the 
$E$- and $B$-mode polarization power spectra. We begin by noting again
that the $T^{\ell}_{m\pm2}$ of equation~\eqref{eq:Tlmk} are closely
related to the spin-2 harmonic coefficients $a_{\pm2 \ell m}$ of $Q$
and $U$ maps constructed from the same TOD. In analogy with
equations~\eqref{eq:almpm2toE} and \eqref{eq:almpm2toB}, we can
therefore define the following $E$-mode-like and $B$-mode-like
linear combinations and two-point correlations of the
$T^{\ell}_{m\pm2}$:
%As the $T^{\ell}_{m\pm2}$ are related to the $a_{\pm2 \ell m}$ from a
%map made using the TOD, the $\tilde{C}_{\ell}^{EE}$,
%$\tilde{C}_{\ell}^{BB}$ etc. defined in \citet{2005MNRAS.360.1262B}
%are related to the $\tilde{C}_{\ell}^{k k'}$. Therefore we can examine
%the effect the asymmetry of the beam has on the measured E- and B-mode
%power spectra if it was not accounted for by defining,
\ba
T_{mE}^{\ell} &\equiv& -\frac{1}{2}\left(T_{m2}^{\ell}+T_{m-2}^{\ell} \right) \label{eq:TlmE}\\
T_{mB}^{\ell} &\equiv& -\frac{1}{2i}\left(T^{\ell}_{m2}-T^{\ell}_{m-2}\right)\:\: \label{eq:TlmB} \text{and}\\
\implies \tilde{C}_{\ell}^{EE} &=&
\frac{1}{4}\left(\tilde{C}_{\ell}^{22} + \tilde{C}_{\ell}^{-2-2} 
+ \tilde{C}_{\ell}^{-22} + \tilde{C}_{\ell}^{2-2}\right) \\
\implies \tilde{C}_{\ell}^{BB} &=&
-\frac{1}{4}\left(\tilde{C}_{\ell}^{22} + 
\tilde{C}_{\ell}^{-2-2} - \tilde{C}_{\ell}^{-22} - \tilde{C}_{\ell}^{2-2}\right)
\ea
%Therefore we examine the effect of the asymmetry and cross polar
%response of the beam by calculating the appropriate coupling
%operators.
In Appendix~\ref{sec:rel_to_stand}, we show that the quantities
defined in equations~\eqref{eq:TlmE} and \eqref{eq:TlmB} are similar to
the standard pseudo-$C_{\ell}$ $E$- and $B$-modes defined in
\citet{2005MNRAS.360.1262B}. These relations can then be used to
investigate the impact of beam asymmetries and/or a non-zero
cross-polar beam response function.
%\subsection{Coupling $C_{\ell}^{TT}$ to $C_{\ell}^{BB}$ due to the asymmetry of the beam} \label{sec:TtoPleak}

Of particular concern is the potential coupling between the
temperature power spectrum and the $B$-mode polarization power
spectrum. {\color{black}As the temperature power spectrum is known to be at least
four orders of magnitude stronger than the $B$-mode power, any
coupling between the two could be catastrophic if not properly
accounted for.} Here we show how the most prominent asymmetric modes of
the beam can potentially create such a coupling.

As described in Section~\ref{sec:reducesum} (see
Figs.~\ref{fig:beam1} and \ref{fig:gaus_decomp}), the $k=\pm2$ mode
is by far the most prominent asymmetric term for the representative
beams that we consider later in this paper. To examine the impact of
the $k = \pm2$ asymmetry, we consider a situation where all
possible orientations are observed, i.e.~where
$W(\bom)n(\bom){=}1$, {\color{black} for the normalisation described in section \ref{sec:setupprob}.}
 In this case, the contribution to
$T_{m\pm2}^{\ell*}$ from the temperature fluctuations is
\ba
T_{m\pm2}^{\ell*} &=& \!\!\!\sum_{\ell' m' k'} b^*_{0 \ell' k'}\,a_{0 \ell' m'} \nonumber \int \vect{d}^3\bom D^{\ell}_{m \pm2} (\bomega)D^{\ell'*}_{m' k'}(\bomega)\nonumber \\
&=& \frac{8\pi^2}{2\ell+1} b^*_{0 \ell \pm 2}\,a_{0 \ell m}, \label{eq:T2P}
\ea
where we have used the orthogonality of the Wigner D-matrices
\citep{1967JMP.....8.2155G}. Equation~\eqref{eq:T2P} shows that the
asymmetry will couple temperautre to polarization even in the case of
an ideal scan strategy. This was to be expected, since both the
CMB polarisation field and the convolution of the temperature
fluctuations with the $k=\pm2$ term of the beam are spin-$\pm$2
quantities. From equations~\eqref{eq:TlmE} and \eqref{eq:TlmB}, the
effect on the measured the $E$- and $B$-mode polarisation is
\ba
\Delta T^{\ell}_{mE}&=& -\frac{4\pi^2}{2\ell+1} (b_{0 \ell 2}+b_{0 \ell -2})a^*_{0 \ell m}\:\:\:\text{and}\label{eq:T2E} \\
\Delta T^{\ell}_{mB}&=& -\frac{4\pi^2}{i(2\ell+1)} (b_{0 \ell 2}-b_{0 \ell -2})a^*_{0 \ell m}. \label{eq:T2B}
\ea
For the coupling from temperature to $B$-mode polarisation to be
non-zero in this case, then $b_{0 \ell \pm2}$ must be complex. This will
only be the case if the beam asymmetry is orientated at an angle to
the polarization sensitivity direction defined by the co-polar
response, which will not be the case in general.

{\color{black} This result was first found by \citet{2003PhRvD..67d3004H}, where they examined 
statistically varying systematic errors. A statistically varying differential beam ellipticity
 between a detector pair would couple temperature to polarisation just
as our constant ellipticity has.
\citet{2007MNRAS.376.1767O} also studied the systematic errors induced when
an elliptical Gaussian beam is used in a $B$-mode experiment. They
considered a specific type of ellipticity: one where the beam is
either perturbed along the direction of the polarisation, or
perpendicular to it. This type of perturbation has the unique
property of having a decomposition where $b_{0 \ell \pm2}$ is real,
and therefore such an asymmetry cannot couple
temperature fluctuations to $B$-mode fluctuations. As \citet{2007MNRAS.376.1767O} shows, 
if the beam is perturbed in any other way than this special
case, then temperature fluctuations will be coupled to $B$-mode
fluctuations even in the case of an ideal scan strategy. This result
is in agreement with the findings of \citet{2008PhRvD..77h3003S} who
also considered the coupling between temperature and $B$-mode
polarisation due to beam asymmetry effects. }

\section{Beam asymmetry correction during map-making}\label{sec:map_making}
In some cases, the approach to correcting for beam asymmetry presented
in Section~\ref{sec:pseudo-cl-technique} will be sub-optimal. For
example, in the case where one corrects for significant
temperature-to-polarization leakage, there will be a contribution to
the error-bars on the reconstructed polarization power spectra due to
the sample variance associated with the leaked temperature signal. In
addition, for polarization experiments that do not measure $Q$ and $U$
simultaneously in the time-stream, the $B$-mode power spectrum errors
will be affected by the sample variance associated with the much
larger $E$-mode polarisation. A technique that corrects for beam
asymmetry at the map level will be immune to these issues and is
therefore an attractive prospect. Here, we develop such a method to
correct for beam asymmetry effects during the map-making step.

We begin by recalling that the convolution of the sky signal with a
general beam is given by equation~\eqref{eq:t_conv}. In an experiment
a telescope will scan the sky, giving us a set of measurements of this
function $t(\bomega)$. For each pixel on the sky we therefore have a
set of measurements at various orientations.

\subsection{Extracting the temperature and polarisation of a pixel} \label{sec:ext_spin_pix}
This discussion is concerned with estimating the temperature and
polarisation signal at the position of a single sky pixel. The
detected signal at the position of a pixel $S$ will depend on the
instrument orientation at the time of observation $\psi$ due to the
polarisation of the sky and the beam asymmetry. If the beam was
axisymmtric and had no polarised response then $S(\psi)$ would be
constant and equal to the beam-smoothed CMB temperature at the pixel
location. \footnote{{\color{black} The dectected signal for a pixel $i$ is related to our 
previous notation by $S_i(\psi)=t(\psi, \theta{=}\theta_i, \phi{=}\phi_i)$  }}
 This can be seen from equation~\eqref{eq:t_conv} by setting
$b_{s\ell k}{=}b_{0\ell 0}\delta_{k0}\delta_{s0}$ and noticing that
the $\psi$ dependence of $t(\bomega)$ comes from the Wigner D
matrices. For a polarised detector $S(\psi)$ would then contain
$k{=}{\pm}2$ Fourier modes, due to the spin-2 nature of
polarisation. This property is exploited by map-making algorithms to
find the temperature and polarisation of a pixel. In these algorithms
the $\psi$ dependence of the detected signal is assumed to be due to the
polarisation signal. Here, we relax this assumption and develop a map-making
algorithm that provides estimates of the temperature and polarisation of
a pixel that are free of systematics of different spins.

We represent the orientations at which a pixel is seen in an
experiment by defining a window function $h(\psi){\equiv}\frac{1}{n_{\rm hits}}\sum_j
\delta(\psi-\psi_j)$, where $n_{\rm hits}$ is the number of hits on the pixel. $h(\psi)$ will be different for
each pixel and will be dependant on the scan strategy. The detected signal $S^d(\psi)$ is therefore
\begin{eqnarray}
S^d(\psi) = h(\psi)S(\psi).
\end{eqnarray}
In Fourier space\footnote{%
We define the Fourier transform of $f(\psi)$ and the inverse transform
to be
$\tilde{f}_k = \frac{1}{2\pi} \int_0^{2\pi} d\psi e^{-ik\psi}f(\psi)$
and $f(\psi) =  \sum_{k{=}-\infty}^\infty e^{ik\psi}\tilde{f}_k$.}
this multiplication takes the form of a convolution
\begin{eqnarray}
%\tilde{S}^d_k &=& \sum_{k'{=}-\infty}^\infty\tilde{h}_{k-k'}\tilde{S}_{k'}\\
\tilde{S}^d_k &=& \sum_{k'{=}-\infty}^\infty H_{kk'}\tilde{S}_{k'},\label{eq:Hbig}
\end{eqnarray}
where we have defined $H_{kk'}{\equiv}\tilde{h}_{k-k'}$. Therefore, if
we can invert the matrix $H_{kk'}$ then we can recover the true
$\tilde{S}_k$. Recovering the spin-0 and spin-2
features of $S(\psi)$ is the main goal of this work since these are the
CMB temperature and polarisation of the pixel. Therefore, we would like to
obtain estimates for $\tilde{S}_0$ and $\tilde{S}_{\pm2}$.

Inverting the matrix $H_{kk'}$ as it is written in equation
\eqref{eq:Hbig} would be impossible: firstly it is infinitely large,
and secondly for any realistic $h(\psi)$ the matrix will be singular.\footnote{% 
Note that ``realistic'' in this context explicitly excludes the case of an
ideal scan strategy for which $h(\psi){=}1$.}
However, if we make the assumption that $\tilde{S}_{k}$ cuts off at a small value of $k_{\rm
  max}$ and if the $\psi$ angle coverage of that pixel is sufficent such that this reduced
 matrix is invertible then we can use the approximation,
\begin{eqnarray}
\tilde{S}_k &=& \sum_{k'{=}-k_{\rm max}}^{k_{\rm max}} H^{-1}_{kk'}\tilde{S}^d_{k'}. \label{eq:spin_sep}
\end{eqnarray}
We can choose $k_{\rm max}$ by measuring the azimuthal dependence of
the beam and ensuring that we include enough $k$-modes to capture all
of the Fourier modes in $S(\psi)$. As in the case of the
pseudo-$C_\ell$ approach (Section~\ref{sec:pseudo-cl-technique}), the
$k_{\rm max}$ should chosen so that the asymmetry of the beam is fully
captured. We return to this issue in the following section where we
examine the harmonic decomposition of some representative asymmetric
beams.

Adopting the {\sevensize
  HEALPix}\footnote{See http://healpix.sourceforge.net} definition of the
Stokes parameters, we can calculate the temperature and polarisation
of the pixel from the inverse Fourier transform of the estimated
$\tilde{S}_k$ as
\ba
T &=& \tilde{S}_0,\\
Q &=& 2\Re(\tilde{S}_2),\\
U &=& 2\Im(\tilde{S}_2).
\ea
{\color{black} Performing this procedure for all observed pixels, we will then have
estimates of the $T, Q$ and $U$ maps which are free of systematics that have
a different spin to our desired quantity. Note we will have not removed systematics
that have the same spin. Specifically the coupling from 
temperature to polarisation due to the asymmetry in the beam discussed in Section
\ref{sec:TtoPleak} will still contaminate our estimates of $Q$ and $U$. We will consider this 
problem in the following subsection. As we show in the simulation tests in Section 
\ref{sec:sim_test_map} 
there will be a noise penalty associated with this effective re-weighting of the data.
For this reason $k_{\rm max}$ should be as large as it needs to be to capture
all the asymmetry in the beam but no larger as this will increase the noise in the map.

This method is similar to the re-weighting of the data to remove systematics of different 
spin presented in \citet{2009arXiv0906.1188B}. Note however that in \citet{2009arXiv0906.1188B}
study suggests descarding (or down weighting) all data that does not have a "counterpart"
that could be used to average systematics to zero. For example, a
spin-1 systematic can be removed if every TOD element has a counter part where the pixel
was hit with $\psi$ orientation $\pi$ away from the first. Conversely, the method we propose
in this paper can use all of the data to characterise the spin-1 systematic and remove it.}

\subsection{Removing the leakage from temperature to polarisation} \label{sec:TtoP_rm}
We showed in Section~\ref{sec:TtoPleak} that the asymmetry of the beam will
leak temperature fluctuations to polarisation regardless of the scan
strategy. This is due to the spin-2 dependence of the temperature of
the CMB convolved with the $k{=}\pm2$ mode of the beam response
function. The polarisation maps made using the map-making method described above
will still contain this spin-2 leakage from temperature to
polarisation. However, since we know that this leakage is from the
temperature fluctuations coupling to the $k{=}\pm2$ asymmetry in the
beam, and since we also now have an unbiased estimate of the
temperature map, we can therefore calculate and remove this leakage.

We start from the estimated temperature map. If we have observed only part of
the sky it is necessary to apodise the map such that there are no
features due to the mask that are smaller than the beam scale. We then
take the spherical harmonic transform of this map,
\ba
\tilde{a}_{0\ell m} &=& \int \: d\Omega\, _0Y^*_{\ell m}(\Omega) \tilde{T}(\Omega).
\ea
We can now calculate the polarization leakage using equations
\eqref{eq:T2E} and \eqref{eq:T2B}. One can show that the spherical
harmonic transform of the temperature leakage is given by

\ba
\Delta a^{E}_{\ell m} &=& 2 \Re\left(\frac{b_{0 \ell 2}}{b_{0 \ell 0}}\right)\tilde{a}_{0\ell m},\\ \label{eq:TtoE_rm}
\Delta a^{B}_{\ell m} &=& 2i \Im\left(\frac{b_{0 \ell 2}}{b_{0 \ell 0}}\right)\tilde{a}_{0\ell m}. \label{eq:TtoB_rm}
\ea
Upon transforming back to real space, these terms can then be removed
from the polarisation maps. Note that the regions within a beam-scale
of the locations where the temperature map was aposided should be
disregarded as the leakage will not have been removed in these regions.

\begin{figure*}
\begin{center}
%\begin{tabular}{c c c}
~\\
~\\
\includegraphics[width=0.3\linewidth, trim=0cm 0cm 0cm 0cm, clip=true]{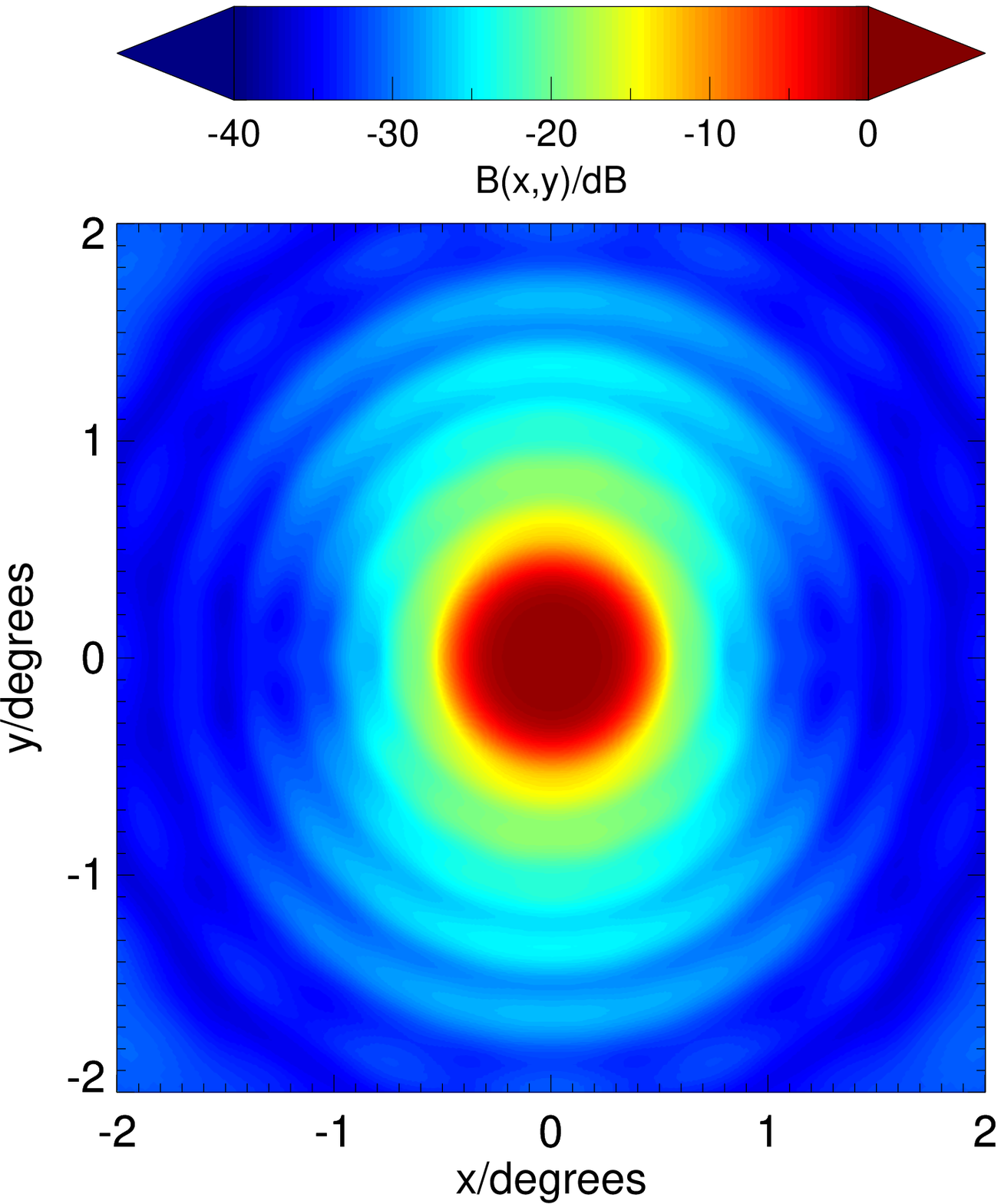}  
\includegraphics[width=0.3\linewidth, trim=0cm 0cm 0cm 0cm, clip=true]{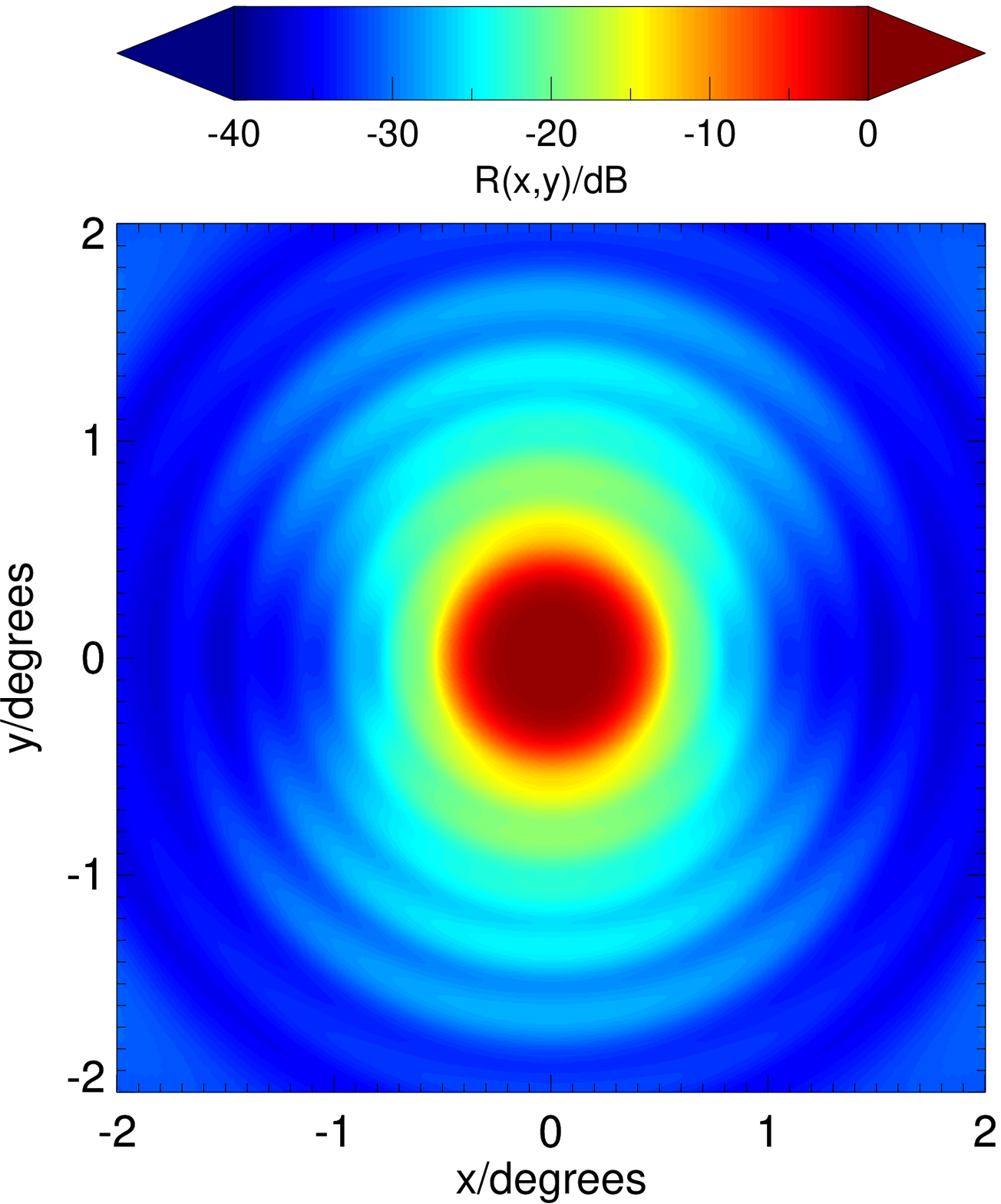}
\includegraphics[width=0.3\linewidth, trim=0cm 0cm 0cm 0cm, clip=true]{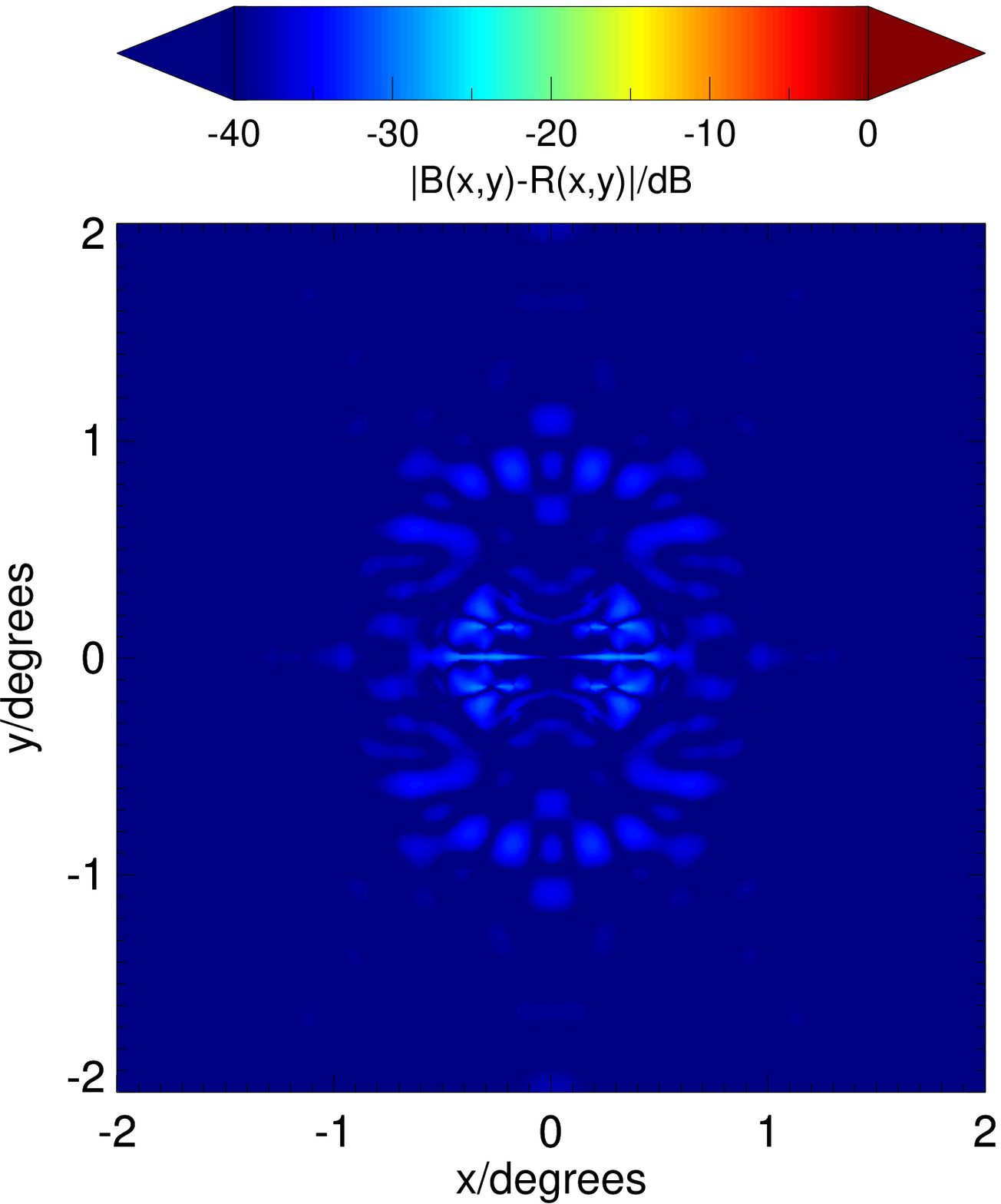}\\

%\end{tabular}
\caption{{\it Left panel:} The simulated beam of a 17 moded horn planned to
  be on board the LSPE balloon experiment. {\color{black}{\it Centre panel:} the
  reconstruction of the beam retaining only modes with $k{=}0,\pm2$ }.
 {\it Right panel:} The absolute error between the
  simulated beam and the reconstruction. \label{fig:beam1}}
\end{center}
\end{figure*}

%\begin{figure*}
%\begin{center}
%\begin{tabular}{c c}
%
%\includegraphics[width=0.45\textwidth]{fig/cg_beam-crop.pdf} & \includegraphics[width=0.45\textwidth]{fig/cg_abs_error-crop.pdf}\\
%
%\includegraphics[width=0.45\textwidth]{fig/cg_recon-crop.pdf} & \includegraphics[width=0.45\textwidth]{fig/cg_frac_error-crop.pdf}\\
%\end{tabular}
%\caption{\textbf{Left:} The simulated beam (top) and the reconstructed beam (bottom) from just 2 modes. $\theta$ and $\phi$ are the usual spherical coordinates. The $\phi$ dependence of the multi-moded horn is evident, but the reconstruction seems to capture 
%the main features. 
%\textbf{Right:} The absolute and fractional errors of the reconstructed beam are plotted (top and bottom respectively) The standard deviation of the residuals $\sigma=9.35\times10^{-5}$ when the peak of the beam is normalized to unity.\label{fig:beam1}}
%
%\end{center}
%\end{figure*}
%\begin{figure}
%\begin{center}
%\begin{tabular}{c}
%
%\includegraphics[width=0.45\textwidth]{fig/cg_abs_error-crop.pdf}\\
%
%\includegraphics[width=0.45\textwidth]{fig/cg_frac_error-crop.pdf}\\
%\end{tabular}
%\caption{The absolute and fractional errors of the reconstructed beam are plotted (top and bottom respectively) The standard deviation of the residuals $\sigma=9.35\times10^{-5}$ when the peak of the beam is normalized to unity.\label{fig:beam2}}
%
%\end{center}
%\end{figure}

\section{Evaluating the parameter \lowercase{$k_{\rm{max}}$} for a
temperature only experiment} \label{sec:reducesum} 

Both the pseudo-$C_\ell$ technique and the map-making approach for
removing beam asymmetries require us to impose a cap on the harmonic
expansion of the beam, $k_{\rm max}$. We now look at a realistic set
up for a CMB experiment in order to ascertain how large the required
$k_{\rm max}$ is likely to be for real experiments. We make no
assumption on the scan strategy. However, beam shapes are generally
designed to be as axisymmetric as possible. We expect therefore that
the beam expansion can be truncated after only a few terms with
minimal loss of accuracy. We note that the pseudo-$C_\ell$ correction
technique can in principle be applied for any $k_{\rm{max}}$ that is
deemed necessary in order to achieve the required accuracy. This will
come with the obvious computational cost of increasing the number
of summations required to evaluate equation~\eqref{eq:oper}. In
contrast, for the map-making approach, there will be some maximum
value of $k_{\rm max}$ for which the matrix $H_{kk'}$ of
equation~\eqref{eq:spin_sep} will be invertible. We also note that increasing $k_{\rm max}$
will increase the statistical error on the recovered map. It should
therefore be chosen to be only as large as is required by the
asymmetry in the beam. In practice, whether the map-making approach
will be appropriate for any given experiment will depend on both the
complexity of the beam asymmetry and on the polarization angle
coverage of the experiment.

\subsection{Beam decompositions for some representative
cases}\label{beam_decomp_examples}
As a demonstration of how a realistic and asymmetric beam can be
represented using just a few $k$-modes, we consider a numerical
simulation of an instrumental beam corresponding to the multi-moded
145~GHz feed horns planned to be flown on-board the balloon-borne
Large Scale Polarisation Explorer
(LSPE, \citealt{2012arXiv1208.0281T}).  By coupling 17 wave-guide
modes, the sensitivity of a single LSPE horn + detector module is
greatly increased. However, the large number of propagated modes
results in a complicated and potentially asymmetric overall beam
shape. Note that the numerical simulation used here is of the horn
only and a simple scaling of the overall beam size has been applied to roughly
approximate the effect of the telescope lens.

We note that the LSPE experiment will also include a half-wave plate
(HWP) in front of the optics which will be used to increase the
polarization angle coverage of the experiment. However, for the
purposes of this demonstration, we consider a temperature-only
experiment and we therefore ignore the effect of the HWP.
%In a future paper, we will present a demonstration of the recovery of
%the polarisation information. 

Upon performing the spin-0 decomposition of the beam,
(equation~\ref{eq:b_decomp_s0}), we find that only a few azimuthal
modes are required to accurately model what might be considered a
rather asymmetric beam. Fig.~\ref{fig:beam1} shows the original
simulated beam alongside a representation of it retaining only the two
most significant modes ($k = 0$ and $k = \pm2$) of the
decomposition. The major features of the beam asymmetry are clearly
well captured using this heavily truncated expansion. We therefore
choose to set $k_{\rm{max}}{=}2$. The beams shown here are invariant
 when rotated by $\pi$, therefore the odd azimuthal terms will be zero.
 Explicitly the truncated decomposition of the beam can be
written as (now writing $b_{0lk} = b_{lk}$ etc. for clarity)
\ba
b_{lk} = b_{l0}\delta_{k0} + b_{l+2}\delta_{k,+2}+b_{l-2}\delta_{k,-2}.\label{eq:beamdec}
\ea
In Fig.~\ref{fig:beam1} we also show the residuals between the full
beam and this representation which is seen to be a good approximation.
We can immediately see from equation~\eqref{eq:cl2pscl} that this will
greatly simplify the pseudo-$C_\ell$ coupling matrix calculation by
limiting the sum over $k_2$ and $k_3$ to only include the $k_2,
k_3{=}0,\pm2$ modes.  

%We note again that there
%is no requirement for the pseudo-$C_\ell$  method to have
%$k_{\rm{max}}{=}2$.  just that it is small as
%computation time and memory requirements will increase rapidly with
%$k_{\rm{max}}$, we use this as a demonstration that our approximation
%is a significant improvement on the axisymmetric approximation.

\begin{figure}
\centering

\includegraphics[width=0.45\textwidth, trim= 1cm 5cm 0cm 0cm]{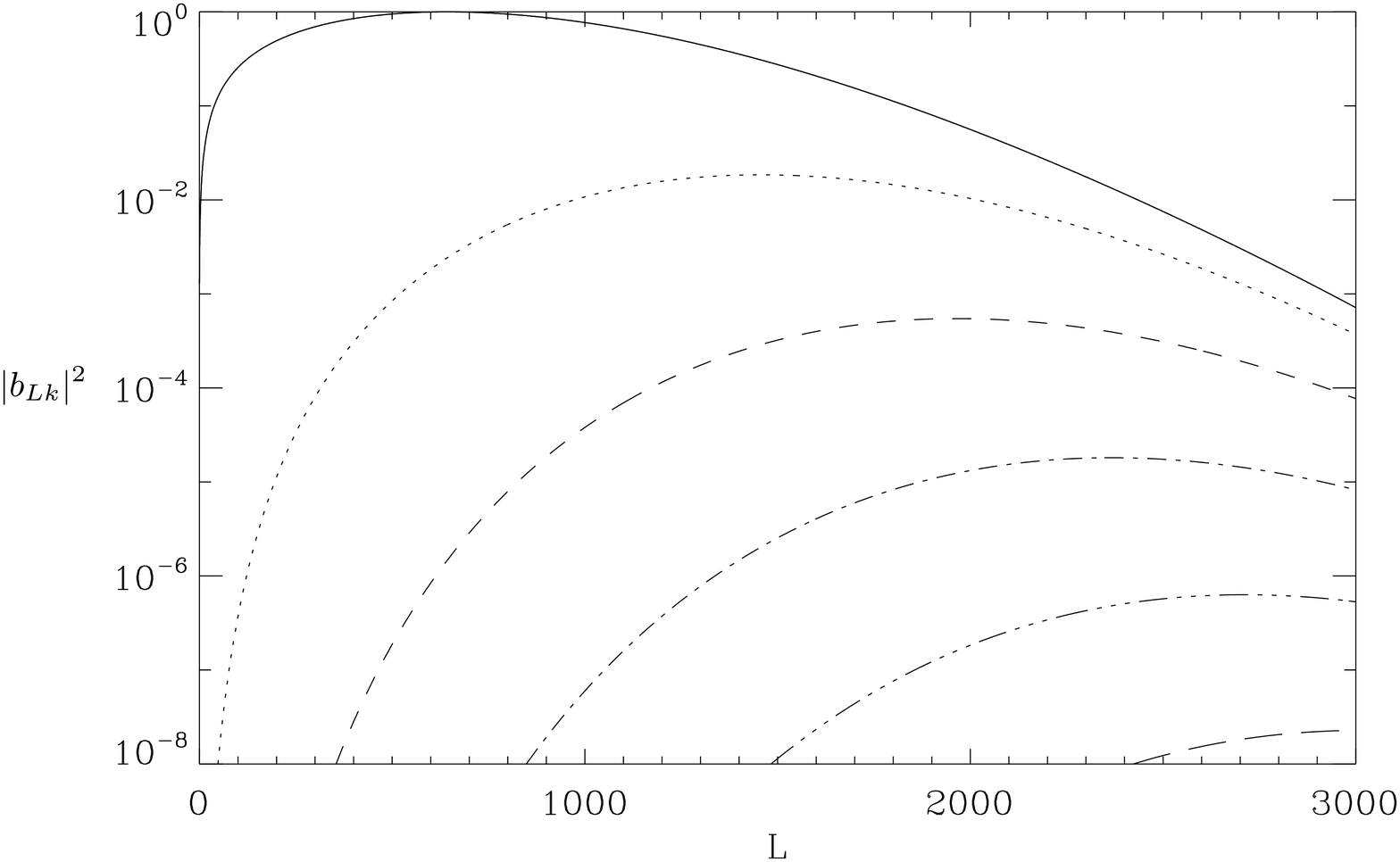}

\caption{The amplitudes of the $k{=}0,2,4,6,8,10$ terms of the beam
  expansion, $|b_{\ell k}|^2$, as a function of multipole, $\ell$ for
  an elliptical Gaussian beam as defined in
  equation~\eqref{eq:gaus_beam} with $\sigma{=}3$ arcmin
  coresponding to a $\theta_{\rm FWHM}{=}7$ arcmin and ellipticity parameter,
  $q{=}1.5$.} \label{fig:gaus_decomp}
%\end{center}
\end{figure}

We also examine a general elliptical Gaussian beam described by 
\ba
B(\theta,\phi) = \frac{1}{2\pi q \sigma^2} e^{-\frac{\theta^2}{2\sigma^2}(\cos^2\phi + q^{-1}\sin^2\phi)}, \label{eq:gaus_beam}
\ea
where $q$ is a parameter that defines the asymmetry of the beam and
the full with at half maximum (FWHM) $\theta_{\rm
FWHM}=2.35\sigma$. If $q{=}1$ then the beam is axisymmetric. To test
the expansion we choose $\theta_{\rm FWHM}{=}7$ arcmin and $q{=}1.5$,
which represents a highly elliptical beam, significantly more elliptical than
one might expect in a real CMB experiment. This simulated beam
therefore provides a stringent test of our correction algorithms. We
plot the spin-0 decomposition of this beam in
Fig.~\ref{fig:gaus_decomp}. It is encouraging to see that once again,
we can describe the beam asymmetry using a relatively small number of
terms.

\subsection{Using the noise power to set $k_{\rm max}$ in a temperature only experiment} \label{sec:set_kmax}
From the above demonstration, it appears that only a small number of
terms in the harmonic expansion of the beam need to be retained in
order to capture the global features of the beam response including
the effects of asymmetry.  

However, to be truly confident that the truncated expansion is an
accurate enough representation of the beam, one would ideally choose
the value of $k_{\rm max}$ such that the any residual error from
mis-representing the beam is smaller (to some tolerance level) than
the statistical error in an experiment. For a particular
asymmetric expansion term 
the leading order contribution to $\tilde{C}_{\ell}^{00}$ is ${\sim}
|\tilde{b}_{0 \ell k}|^2 C_{\ell}^{00}$, where $\tilde{b}_{0 \ell k} {=} b_{0 \ell
k}\sqrt{\frac{4\pi}{2\ell+1}}$. Therefore we can say that the
asymmetric term should be included if
\ba
\langle p_{k}\rangle_{\rm pix} |\tilde{b}_{0 \ell k}|^2 C_{\ell}^{00} > f\langle N_{\ell}^{00}\rangle_{\sigma},
\label{eq:kmax_noise_rule}
\ea
where $f$ is a tolerance level to which we require our systematic
errors to be below the statistical error. A reasonable choice may be
$f{=}0.1$, to ensure that the systematic error is 10\% of the noise
statistical error. {\color{black}$\langle N_{\ell}^{00}\rangle_{\sigma}$ is the
standard deviation of the noise power, see equation \eqref{eq:noise_power},
which is the dominating source
of statistical error at high $\ell$.} The term $\langle
p_{k}\rangle_{\rm pix}$ is the $k^{\rm th}$ mode of the $\psi$
angle coverage quality of the scan strategy defined as
\ba
\langle p_k \rangle_{\rm pix} = \left< \frac{1}{n^2_{\rm hits}}\left| \sum_i e^{ik\psi_i} \right|^2\right>_{\rm pix}, \label{eq:p_qual}
\ea
where $\langle \rangle_{\rm pix}$ denotes an average over all pixels on
the sky, $n_{\rm hits}$ is the number of hits a pixel receives and the
sum over the index $i$ is over all observations of a particular pixel. The
lower the value of $\langle p_k \rangle_{\rm pix}$, the better the
scan strategy is at removing the bias created by the beam asymmetry
due to the $k^{\rm th}$ azimuthal mode.

The rule of thumb in equation~\eqref{eq:kmax_noise_rule} becomes
problematic when the sky coverage is small and in the presence of a
highly asymmetric beam due to coupling between asymmetric terms of
different $k$. This coupling is suppressed in the all-sky
case because the cross spectra $\mathcal{W}_{k_2 k_3}^{\ell}$ are
small for $k_2{\neq} k_3$, which in turn means the coupling matrix
$M_{00k_2 k_3}^{\ell_1 \ell_2}$ is small.

However in the case where the power in the mask extends to high $\ell$
these cross spectra terms can become significant. This results in a
significant contribution to the operator by a term of the form
$b_{0 \ell_2 k_2}^* b_{0 \ell_2 k_3} M_{00k_2 k_3}^{\ell_1 \ell_2}$,
where $k_2{\neq} k_3$. This additional contribution could potentially
be larger than the $k_2 {=} k_3$ term if $|b_{0 \ell k_2}| {\gg}
|b_{0 \ell k_3}|$.

\section{Testing the pseudo-$C_\ell$ on Simulations}\label{sec:sim_test}
In this section, we demonstrate the implementation of the
pseudo-$C_\ell$ technique developed in
Section~\ref{sec:pseudo-cl-technique} on numerical simulations of two
representative types of CMB experiment -- a generic balloon-borne
experiment and a future satellite mission. Note that we test the
performance of the algorithm in the temperature-only case. The
implementation of our technique for polarization experiments will be
presented in a future paper. For a temperature-only analysis,
the coupling is reduced to 
\ba
\tilde{C}^{00}_{\ell_1} &=& \sum_{\ell_2} O^{0000}_{\ell_1 \ell_2} C^{00}_{\ell_2} \:\: \text{with,} \label{eq:temp_test_pscl}\\
O^{0000}_{\ell_1 \ell_2} &=& \sum_{k_2 k_3} b_{0 \ell_2 k_2}^* b_{0 \ell_2 k_3} M_{00k_2 k_3}^{\ell_1 \ell_2}\label{eq:temp_test_oper}
\ea
The main stages of the simulation pipeline are as follows:
\begin{enumerate}
\item We use the {\sevensize HEALPix} package to create simulations of
  the CMB temperature field based on the theoretical CMB power
  spectrum for the concordance $\Lambda$CDM cosmological model.  
\item We then create a TOD by convolving the CMB with the telescope
  beam in the appropriate orientation given the scan strategy. 
\item We then calculate the $\tilde{C}_\ell^{00}$ using equation~(\ref{eq:pscl}).
\item The coupling operator is calculated using the expression in equation \eqref{eq:temp_test_oper}.
\item With the operator $O^{0000}_{\ell_1 \ell_2}$ calculated, we then
  invert it to recover an estimate of the full-sky and beam-deconvolved power spectrum.
\end{enumerate}

%In the following subsections we describe the various stages of the code developed to recover the power spectra and to test it on a simulated experiment.
%
%\subsection{Simulating a pseudo-$C_\ell$} \label{sec:pscl_sim}
%
Note that we use two different convolution codes for the two classes
of experiment that we simulate. For the balloon-like test we calculate each
element of the TOD using equation~(\ref{eq:td}). The
{\sevensize HEALPix} {\sevensize ROTATE\textunderscore ALMS} routine
operates on the beam multipole coefficients. This function uses the
Wigner D-matrices to rotate the beam coefficients by a set of Euler
angles given on input. The TOD element can then be calculated using,
\ba 
t_j = \sum_{lm}b^{j*}_{lm}a_{lm}, 
\ea 
where $b_{lm}^{j*}$ is the multipole expansion of the beam when it has
been rotated through the Euler angles $\bomega_j$. In this way the
whole TOD can be created. While this is accurate and simple it would
be impossible to use this approach in the satellite-like experiment as
the number of TOD elements and the required $\ell_{\rm max}$ would
make it computationally infeasible. To create the simulated TOD in the
satellite-like case, we use a method similar to the {\sevensize
  FEBeCoP} approach~\citep{2011ApJS..193....5M} except that do not
calculate the effective beams but rather we produce the individual TOD
elements. The convolution is simplified by describing the beam in
pixel space and we only calculate the value of the beam on a set of
neighbouring pixels centred around the pointing centre of the beam. We
use pixels within $5\sigma q$ of the centre of the beam (see
equation~\ref{eq:gaus_beam}) to describe the beam over this localised
region of sky.

A second technical challenge was to obtain the Wigner D-matrix
transforms of the TOD and the window function. We achieved this by
implementing equations~\eqref{eq:Tlmk} and \eqref{eq:wlmk} using the expression for
the Wigner D-matrix in terms of spin weighted spherical harmonics
(equation~\ref{eq:wig_spin}). First, the integral over $\psi$ was
performed using a simple Newton$\mbox{-}$Cotes numerical integration
method. The {\sevensize HEALPix} routine {\sevensize MAP2ALM\_SPIN} was
then used to perform the integrals over $\theta$ and $\phi$.

\subsection{Balloon-like experiment}\label{sec:LSPE_test}
For our first test we use the simulated LSPE beam shown in
Fig.~\ref{fig:beam1} with a modification. We greatly amplify the size
of the asymmetry in order to produce a more stringent test of the asymmetry
correction algorithm. The spherical harmonic decomposition of this
exaggerated beam is related to the original LSPE decomposition by 
\ba
b^{\rm exagg}_{0 \ell, k} = b^{\rm LSPE}_{0 \ell, 0} \delta_{k0} +
b^{\rm LSPE}_{0 \ell, \pm 2} \delta_{k \pm 2} \times 100. \label{eq:x_beam} 
\ea 

We use the planned scan strategy for LSPE but we stress that we have
not included the effects of the half-wave plate (HWP) that LSPE will
use. The LSPE experiment will incorporate a HWP in front of the optics
which will be used to increase the polarization angle coverage of the
experiment. However, for the purposes of this demonstration where we
consider a temperature-only experiment, we have ignored the
(advantageous) effects of the HWP. LSPE will be launched at a latitude
of 78$^0$ North and will follow a jet stream around the globe staying
at roughly the same latitude. The payload will spin at 3 rpm and the telescope
will perform constant allevation scans with regular (\S daily) steps
in elevation \citep{2012arXiv1208.0281T}.
%In a future paper, we will present a demonstration of the recovery of
%the polarisation information. 

\begin{figure*}
\begin{center}
\begin{tabular}{c c}

\includegraphics[width=0.45\textwidth, angle=0, trim=0cm 1.5cm 0cm 2.7cm, clip=true]{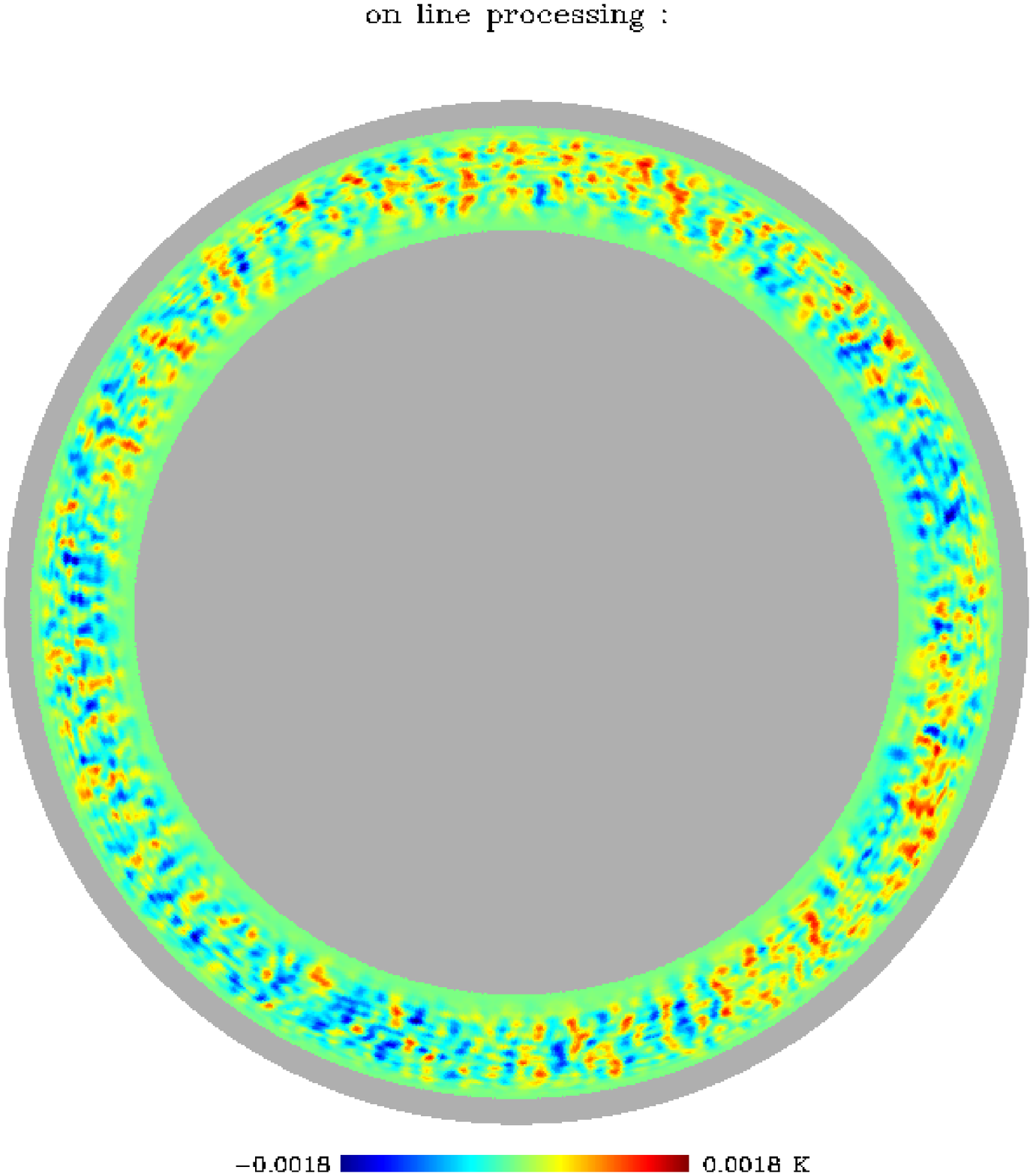}&
\includegraphics[width=0.45\textwidth, angle=0, trim=0cm 1.5cm 0cm 2.7cm, clip=true]{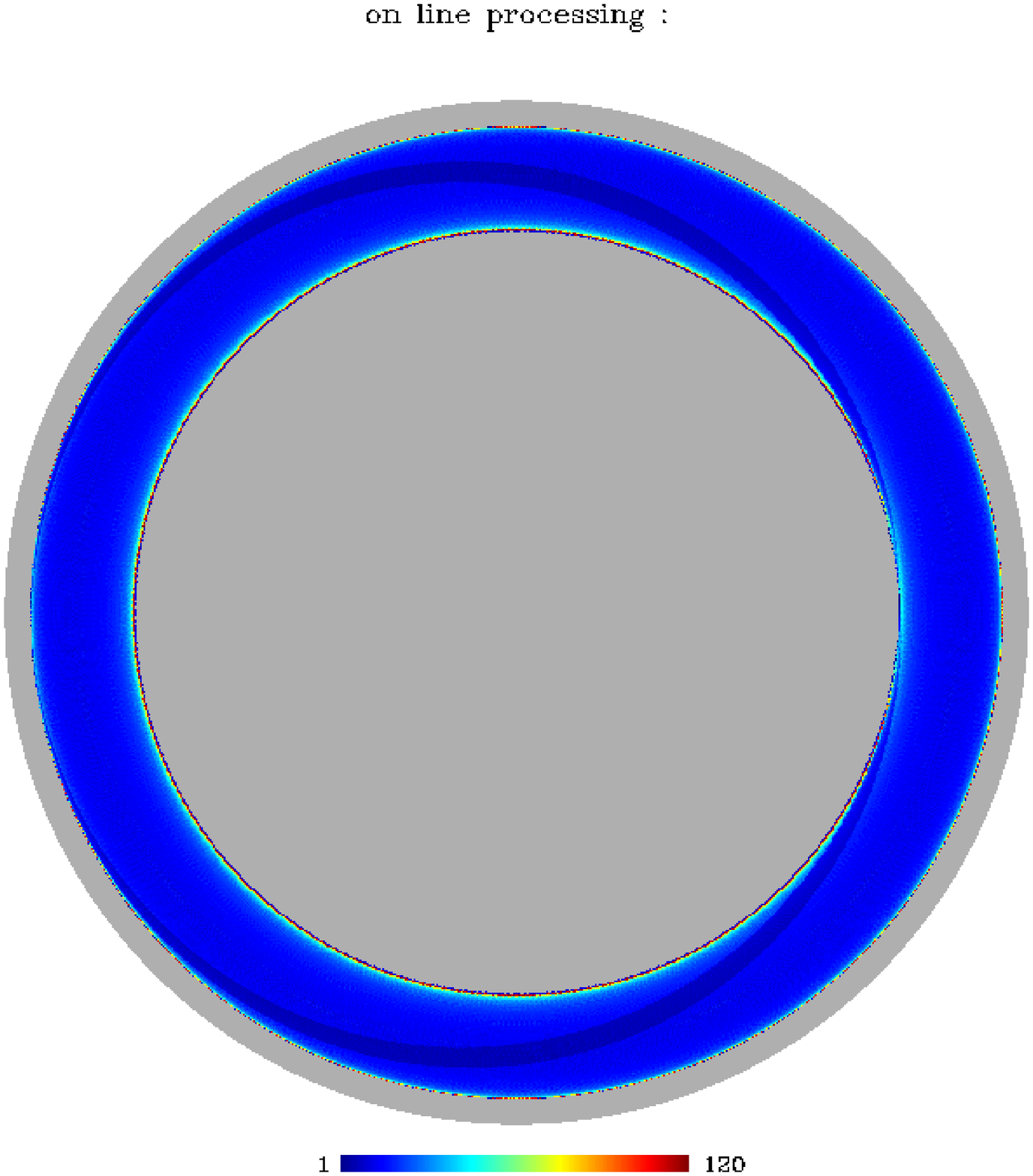}\\
\includegraphics[width=0.45\textwidth, angle=0, trim=0cm 1.5cm 0cm 2.7cm, clip=true]{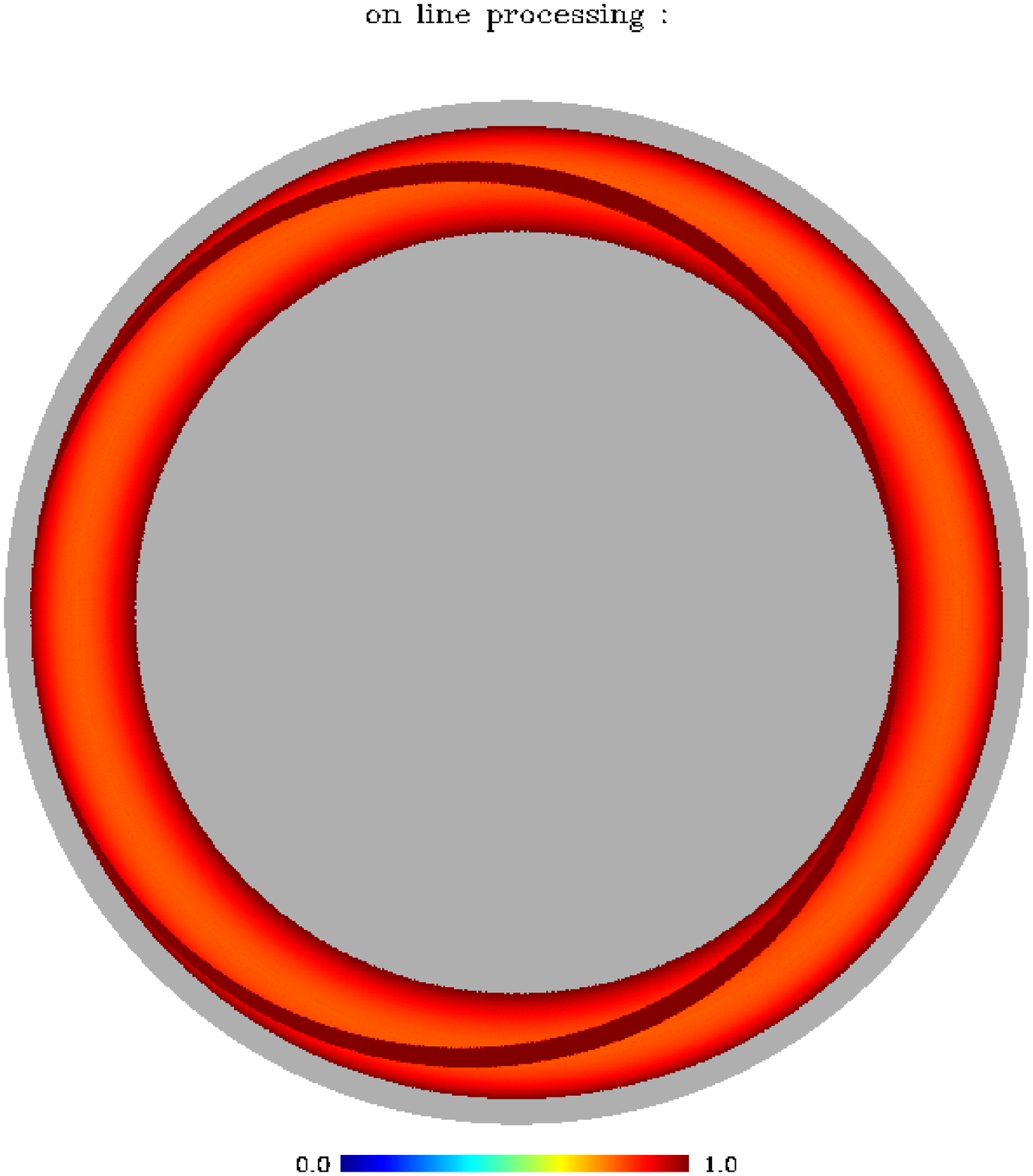}&
\includegraphics[width=0.45\textwidth, angle=0, trim=0cm 1.5cm 0cm 2.7cm, clip=true]{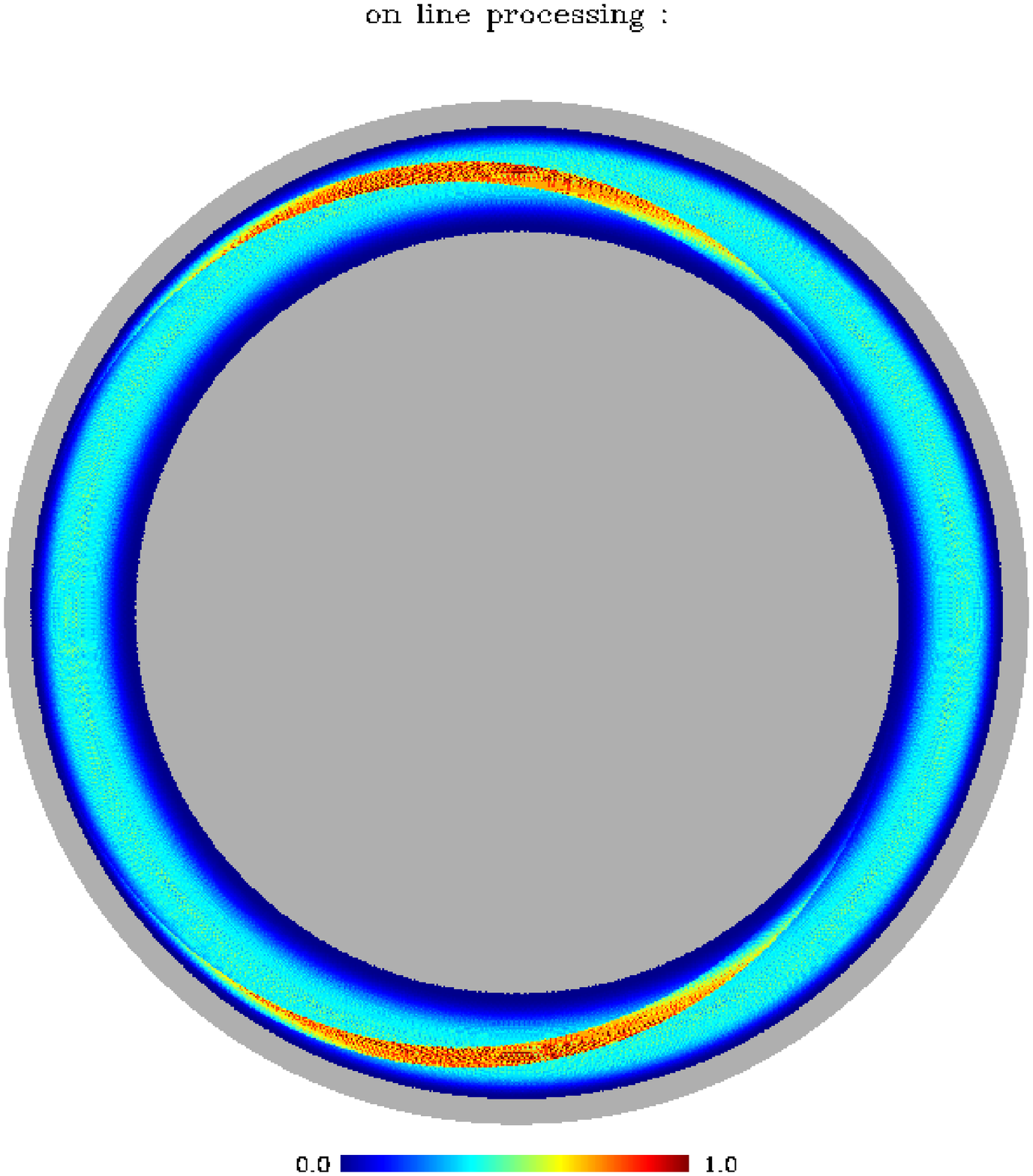}\\
\end{tabular}
\caption{\label{fig:extreme_tod}\textit{Top left}: An example binned
  map made in the balloon-like test. The exaggerated beam was used
  with a spherical harmonic decomposition as described in equation
  \eqref{eq:x_beam}. The scan strategy implemented is typical of a
  balloon based experiment. \textit{Top right}: The hit map from one
  day of observations for the balloon-like experiment. \textit{Bottom
    left}: The $p_2$ quality, defined in equation \eqref{eq:p_qual},
  of the scan strategy. We note again that a HWP was not included
  which is why the polarization angle coverage is
  poor for this simulation. \textit{Bottom right:} The weighting function applied
  to apodise the hit map.}
\end{center}
\end{figure*}

We created a set of TOD following the above procedure including
white noise. The LSPE scan strategy covers $\sim 25\%$ of the sky and
so provides a test of the algorithm's ability to account for a cut sky
in addition to the effects of beam
asymmetry. Fig.~\ref{fig:extreme_tod} shows an example of a simple
binned map created from one of our simulations. This figure also shows
the hit map of the scan strategy, the $2^{\rm nd}$ $\psi$ coverage
quality $p_2$ defined in Section~\ref{sec:set_kmax} and the weighting
function that we use to apodise the hit map. We include an
(unrealistically high) noise level of $1.6$ mK-arcmin in our
simulations in order to test the removal of noise bias from the
recovered power spectrum.

% lspe_example_map.png

\begin{figure}
\begin{center}

\includegraphics[width=0.45\textwidth, trim=1cm 5cm 0cm 0cm]{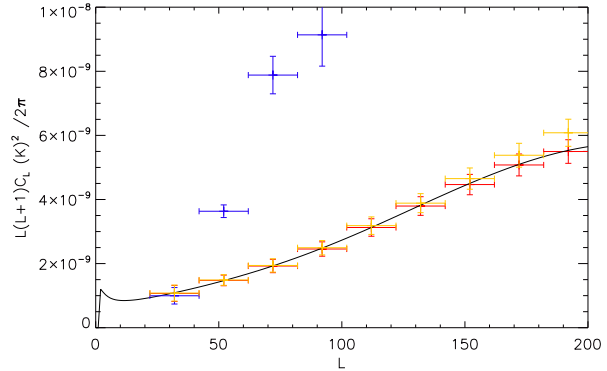}

\caption{Reconstruction of the power spectrum for the balloon-like
  simulation using the pseudo-$C_\ell$ method. The input theoretical
  power spectrum is plotted (black) along with the recovered power
  spectrum when the exagerrated beam is used (red), averaged over 1000
  realisations. We also plot the recovered power spectrum one would
  recover if the axisymmetric MASTER \citep{2002ApJ...567....2H}
  analysis was used (blue) and if the noise was ignored (yellow). The
  vertical error bars show the standard deviation of the realisations
  from the mean.} \label{fig:LSPE_x_cl}

\end{center}
\end{figure}

The $\tilde{C}^{00}_\ell$ for 1000 realisations were computed
according to equation~\eqref{eq:temp_test_pscl} and the coupling
operator was calculated using equation~\eqref{eq:temp_test_oper}. In
order to invert the coupling matrix we needed to bin the
$\tilde{C}^{00}_\ell$ and the coupling operator, as in
\citet{2002ApJ...567....2H}. This binning ensures that the matrix is
invertible and that the resulting data points are uncorrelated. Here
we use a bin size of $\Delta \ell{=}20$. A correction for noise bias
was also implemented following the procedure outlined in
Section~\ref{sec:pseudo_cl_noise}. Fig.~\ref{fig:LSPE_x_cl}
shows the mean power spectrum recovered from the simulations in
comparison to the input spectrum. In this figure, we also plot the
mean power spectrum recovered when the beam was assumed to be
axisymmetric, by including only the $b_{0\ell0}$ term on the pseudo-$C_\ell$.
  This figure clearly demonstrates that our algorithm can
successfully recover the power spectrum in a situation where the
axisymmetric approximation clearly fails. The power spectrum obtained in
the case where the noise in the TOD was ignored is also shown
demonstrating the ability of our algorithm to successfully
correct for noise bias.

\subsection{Satellite-like experiment}\label{sec:epic_test}

\begin{figure*}
\begin{center}
\begin{tabular}{c}

\includegraphics[width=0.6\textwidth, angle=180, trim=0cm 1cm 0cm 0cm, clip=true]{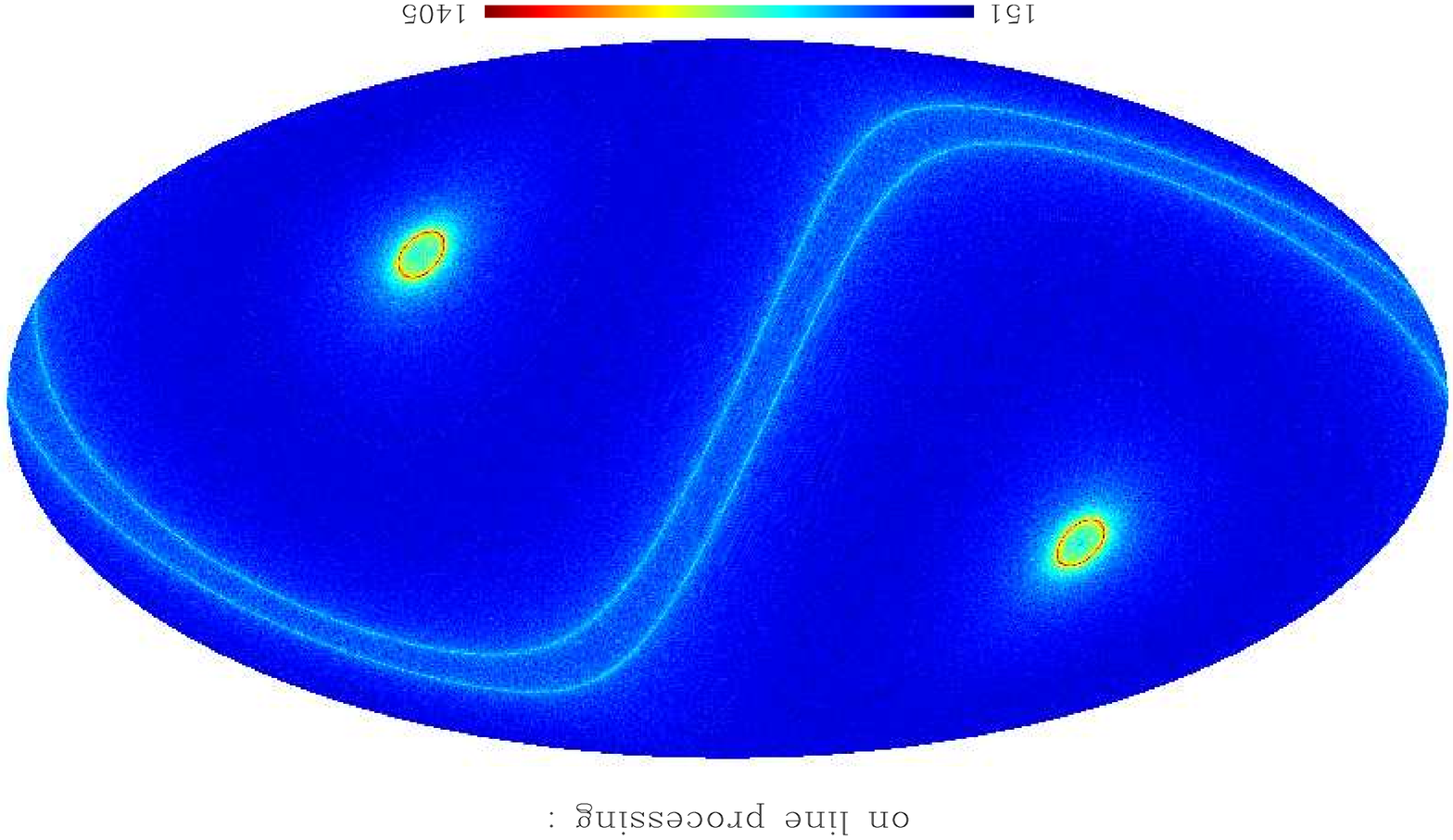}\\
\includegraphics[width=0.6\textwidth, angle=180, trim=0cm 1cm 0cm 0cm, clip=true]{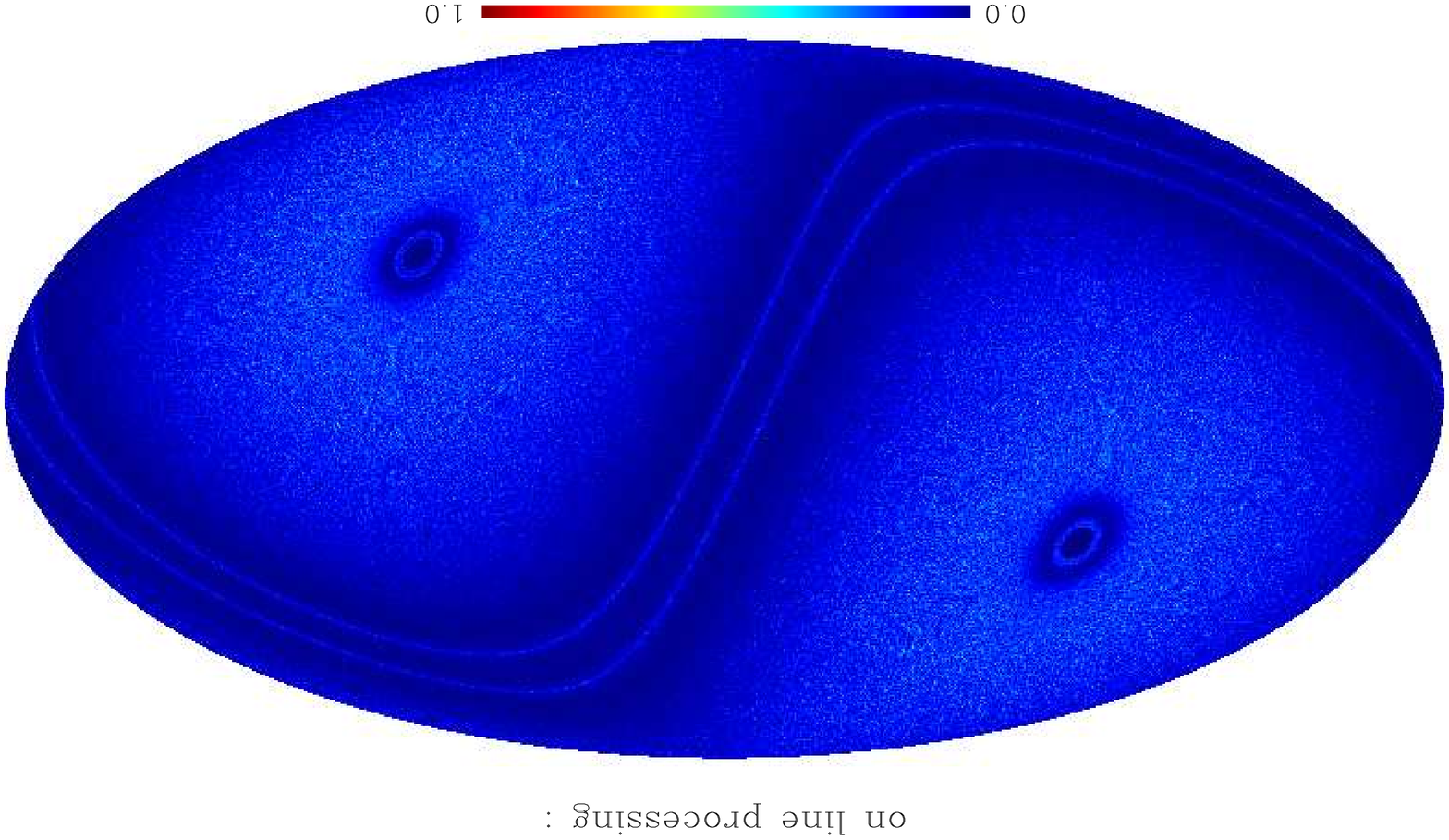}\\
\includegraphics[width=0.6\textwidth, angle=180, trim=0cm 1cm 0cm 0cm, clip=true]{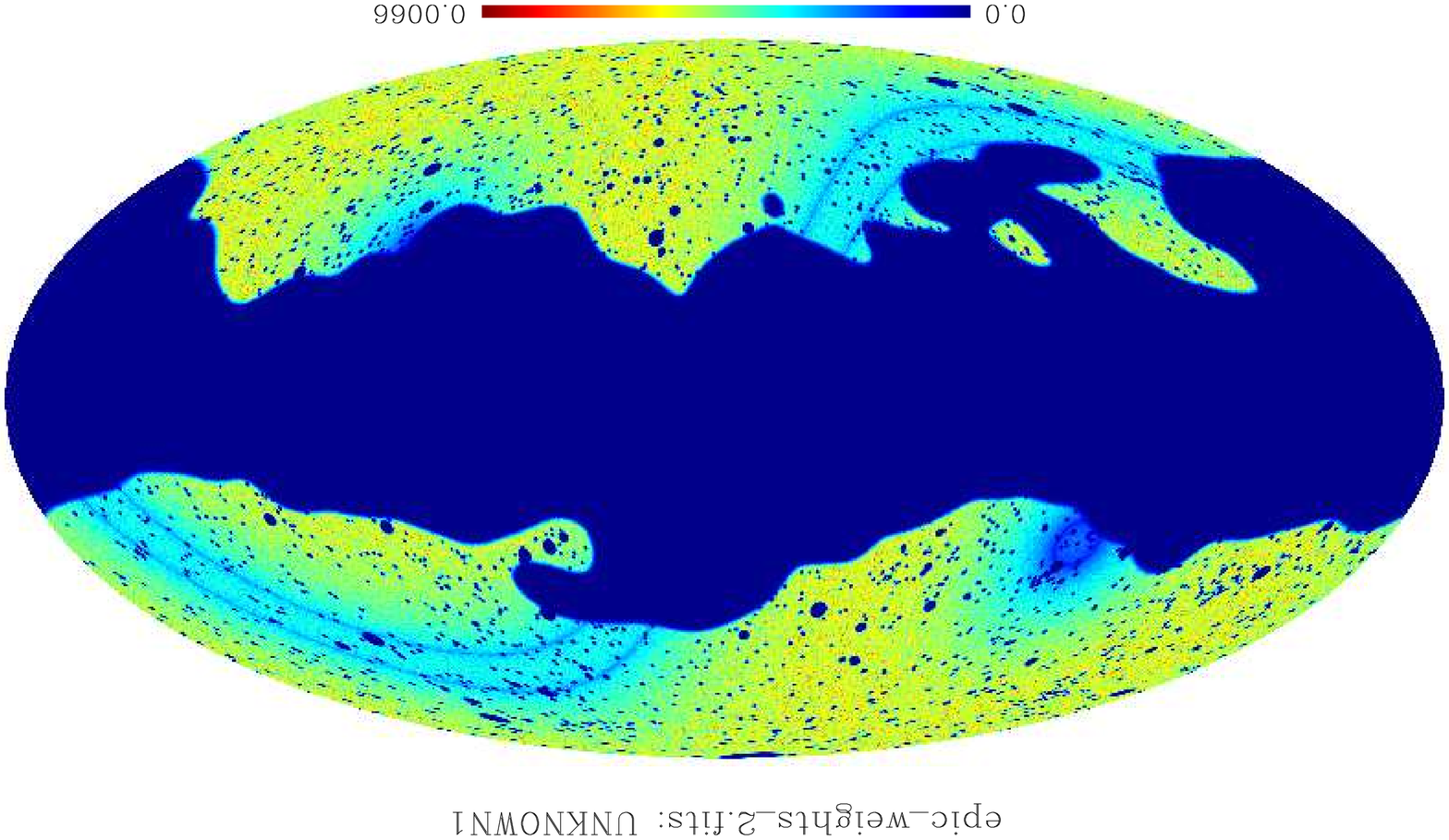}\\
\end{tabular}
\caption{\label{fig:EPIC_scan}\textit{Top panel}: The hit map for one
  year of observations for our simulated satellite-like
  experiment. \textit{Middle:} The $p_2$ quality, defined in equation
  \eqref{eq:p_qual}, of the scan strategy. \textit{Bottom panel:} The
  weighting function used to apply a galactic and point source mask
  and apodise the hit map.}
\end{center}
\end{figure*}

\begin{figure}
\begin{center}
\begin{tabular}{c}

\includegraphics[width=0.45\textwidth, angle=0, trim=0cm 0cm 0cm 0cm, clip=true]{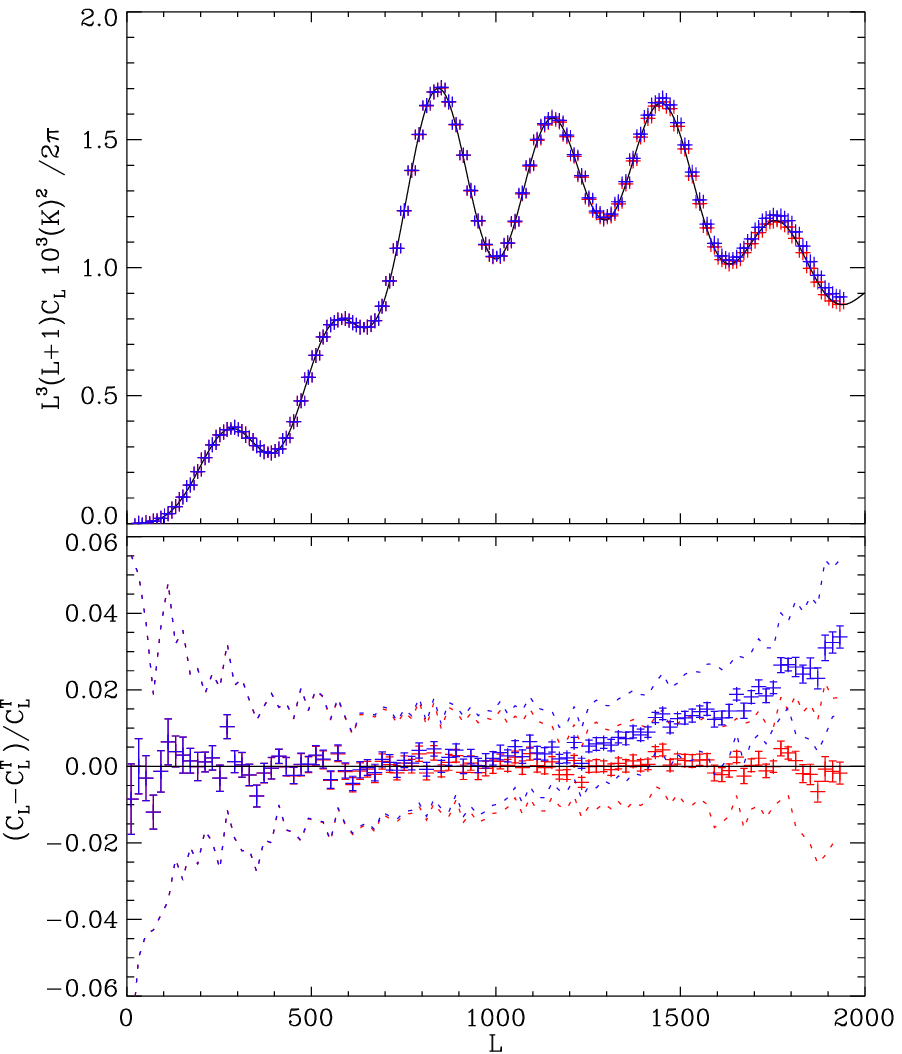}\\
\end{tabular}
\caption{\label{fig:EPIC_cb}Reconstruction of the power spectrum for
  the satellite-like simulation using the pseudo-$C_\ell$ method. {\it
    Top:} The input theoretical power spectrum is plotted (black)
  along with the recovered power spectrum for $k_{\rm max}{=}0,2$
  (blue and red points respectively) averaged over 48 realisations. {\it Bottom:} The
  fractional error between the recover band powers and the binned
  input power spectrum. The error bars show the statistical error on the
  recovered mean, while the dashed lines show the 1$\sigma$ error that
  would be seen in a real experiment. We see that the $k_{\rm
    max}{=}2$ recovery is unbiased to a small fraction of the statistical error
  caused by noise.}

\end{center}
\end{figure}

The second test that we perform is using a satellite-like scan
strategy. For this test, we use a beam described be
equation~\eqref{eq:gaus_beam} with $\sigma{=}3$ arcmin corresponding
to a FWHM of 7 arcmin and $q{=}1.5$. The scan strategy used is similar
to that proposed for the Experimental Probe of Inflationary Cosmology
(EPIC, \citealt{2009arXiv0906.1188B}). In Fig.~\ref{fig:EPIC_scan} we
show the hit map, the $p_2$ quality (equation~\ref{eq:p_qual}) and the
weighting function that we have used to mask the galaxy and
extragalactic sources. We use the same mask as was used by the {\it Planck} collaboration
 \citep{2013arXiv1303.5068P} for their power spectrum analysis. The weighting function also apodises the
hit map. Note that the EPIC scan strategy was designed to improve the
$2^{\rm nd}$ $\psi$ coverage quality defined in
Section~\ref{sec:set_kmax}. This choice of scan strategy will
therefore be effective in mitigating beam asymmetries even in the
absence of a dedicated correction algorithm. We therefore expect the
bias, in the absence of a dedicated correction, to be much smaller
than seen in the previous section (although it will still be
non-zero). The mean recovered power spectra using  $k_{\rm max}{=}0,2$
are shown in the top panel of Fig.~\ref{fig:EPIC_cb}. The lower panel
of this figure shows the fractional residual bias in the power
spectrum recovery and demonstrates an unbiased result to high-$\ell
$ when $k_{\rm max}{=}2$ is used. These simulations included white
noise such that the error on the map is equivalent to
$8.5\,\mu$K-arcmin. We can see that even with a scan strategy designed
to optimise crossing angles and thereby reduce
the asymmetry bias, there remains a bias at the level of $\sim\!\!1{-}2\sigma$ 
when the axisymmetric approximation is made.

The importance of this test is two-fold. It demonstrates that the
algorithm can deal with a more conventional elliptical Gaussian beam
and also that it can be extended to high-$\ell$. The coupling operator
was calculated in 9 hrs on one 2.26 GHz processor to 
$\ell_{\rm max} = 4000$. We calculated the coupling operator to a
much higher $\ell$ than the maximum multipole of interest to ensure
that we did not miss any effects from the aliasing of power from
higher multipoles.

\section{Testing the map-making algorithm on Simulations}\label{sec:sim_test_map}
We test the map-making algorithm described in
Section~\ref{sec:map_making} in two ways. First, we use the algorithm
on multiple realisations of the same simulated experiment as described
in Section~\ref{sec:epic_test} to show that the algorithm can
correctly remove the asymmetry bias on the temperature power
spectrum. The second test is performed on a single noise-free full sky
simulation of the satellite experiment. We use this latter simulation to
test if the algorithm can successfully remove the asymmetry bias
directly from the temperature and polarisation maps. We do not
include a CMB polarisation signal in our TOD and so a successful
asymmetry correction algorithm should recover a zero polarisation
signal.

\subsection{Removing the asymmetry bias on the temperature power spectrum}
As in Section~\ref{sec:epic_test}, we create 48 TOD realisations
using the satellite-like scan strategy, and using an elliptical
Gaussian beam with $\theta_{\rm FWHM}{=}7$ arcmin and $q{=}1.5$ (see
equation~\ref{eq:gaus_beam}). As before, the TOD included white noise
equivalent to a map noise level of $8.5\,\mu$K arcmin.

\begin{figure}
\begin{center}
\begin{tabular}{c}

\includegraphics[width=0.45\textwidth, angle=0, trim=0cm 0cm 0cm 0cm, clip=true]{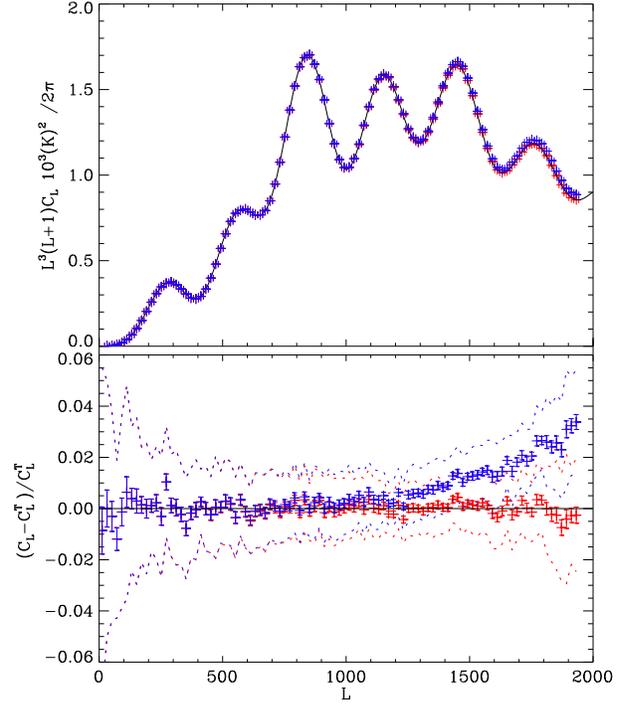}\\
\end{tabular}
\caption{\label{fig:map_cb_plot}Reconstruction of the power spectrum
  for the satellite-like simulation using the map-making method. This 
should be compared to Fig.~\ref{fig:EPIC_cb} which presents the results of the 
pseudo-$C_{\ell}$ method. {\it Top:} The input theoretical power spectrum is plotted (black)
  along with the recovered bandpowers for the binned map
  (blue points), and where the map-making algorithm is used with $k_{\rm
    max}{=}2$ (red points), averaged over 48 realisations. {\it Bottom:} The
  fractional error between the recover bandpowers and the binned
  input power spectrum. The error bars show the statistical error on the
  recovered mean, while the dashed lines show the 1$\sigma$ error that
  would be achieved by a single experiment. We see that the map-making
  technique using $k_{\rm max}{=}2$ produces power spectrum estimates
  that are unbiased to a small fraction of the statistical error
  caused by noise.}
\end{center}
\end{figure}

From the simulated TOD, we then compute both a simple binned map and a
map constructed using the algorithm presented in
Section~\ref{sec:map_making} with $k_{\rm max}{=}2$. We
then apply the same galactic and point source mask to these maps as used in
Section~\ref{sec:epic_test}. A standard pseudo-$C_\ell$
analysis~\citep{2002ApJ...567....2H} is then applied in order to
recover the power spectrum. {\color{black} The recovered power spectrum is 
then deconvolved by dividing by the beam window function $B_{\ell}{=} 
\sqrt{\frac{4\pi}{2\ell+1}}b_{0\ell0}$, which is
the axisymmetric component of the input beam.} Fig.~\ref{fig:map_cb_plot} shows the
recovered power spectrum for the binned map and for the asymmetry
cleaned map obtained using the new map-making algorithm of
Section~\ref{sec:map_making}. The asymmetry bias has clearly been
removed successfully. Note that the removal of the asymmetry bias
comes at the cost of a modest inflation of the power spectrum
error-bars which have increased by ${\sim}20\%$. This increased error
is modest compared to the systematic bias that has been removed.  

\subsection{Removing the asymmetry bias on the temperature and polarisation maps}
Here we present a demonstration of how the map-making algorithm can be
used to remove the asymmetry bias on both the temperature and
polarisation maps. 

\begin{figure*}
\begin{center}
\begin{tabular}{c}

\includegraphics[width=0.55\textwidth, angle=180, trim=0cm 1cm 0cm 0cm, clip=true]{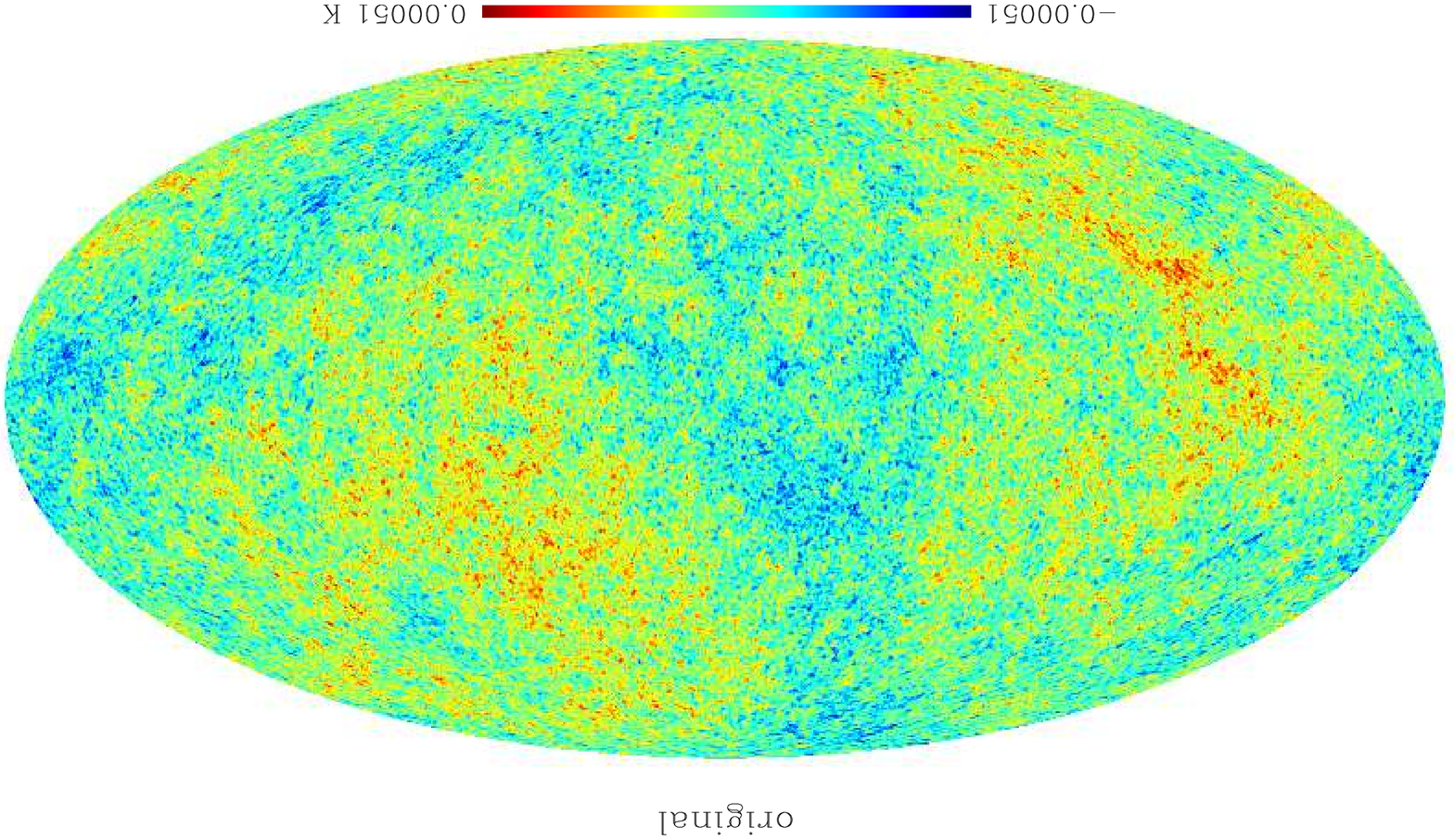}\\
\includegraphics[width=0.55\textwidth, angle=180, trim=0cm 1cm 0cm 0cm, clip=true]{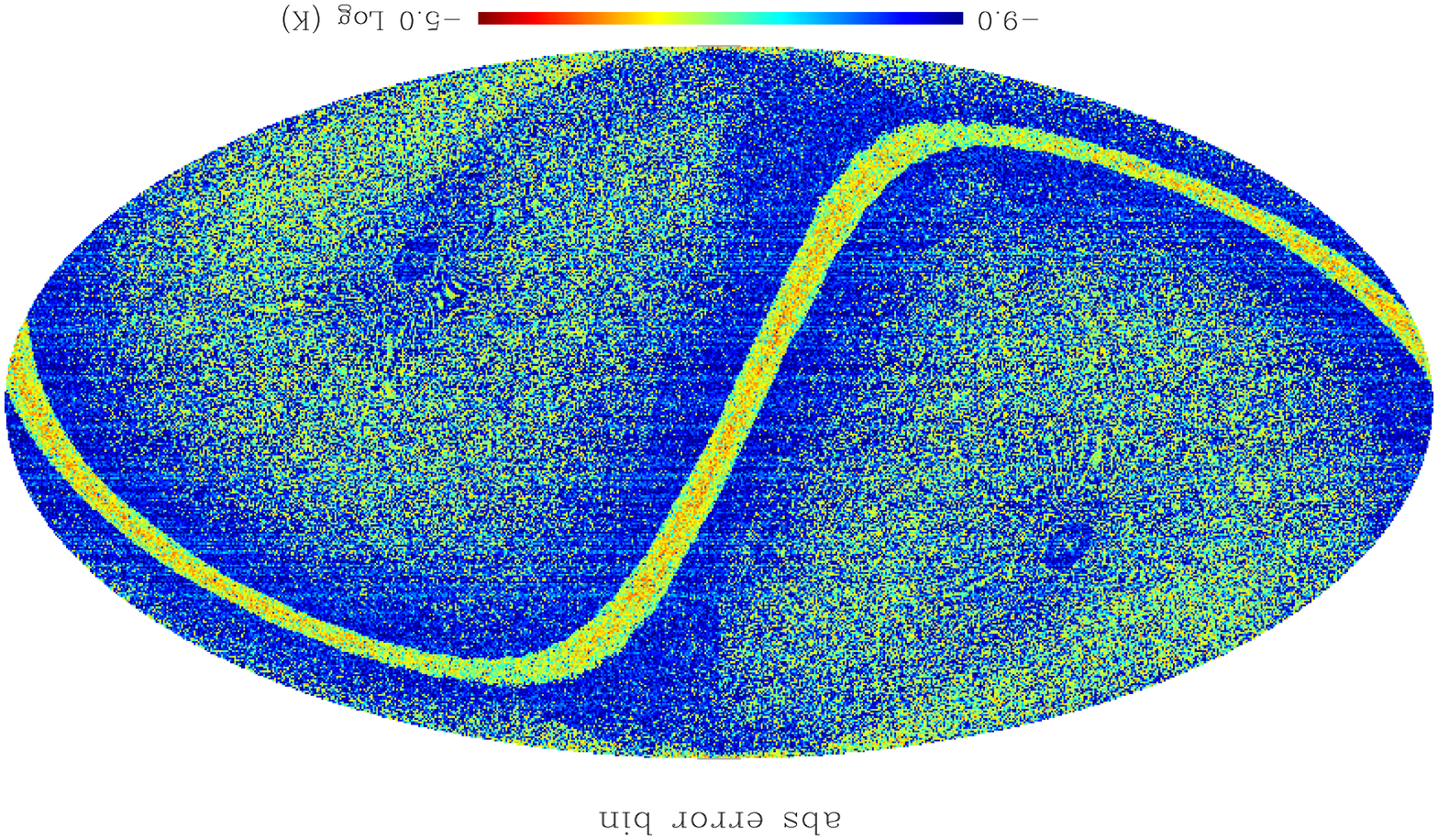}\\
\includegraphics[width=0.55\textwidth, angle=180, trim=0cm 1cm 0cm 0cm, clip=true]{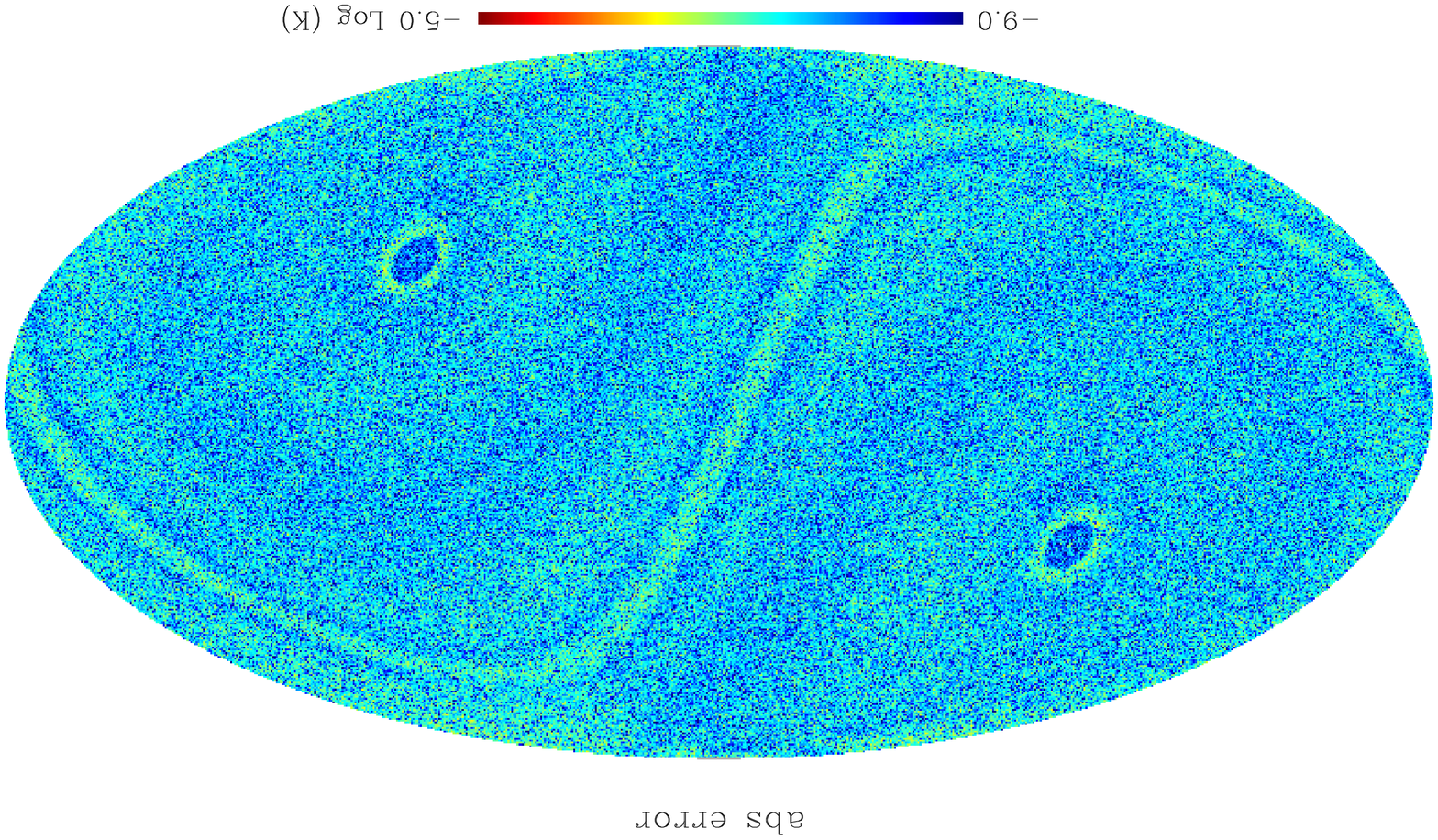}\\
\includegraphics[width=0.55\textwidth, angle=180, trim=0cm 1cm 0cm 0cm, clip=true]{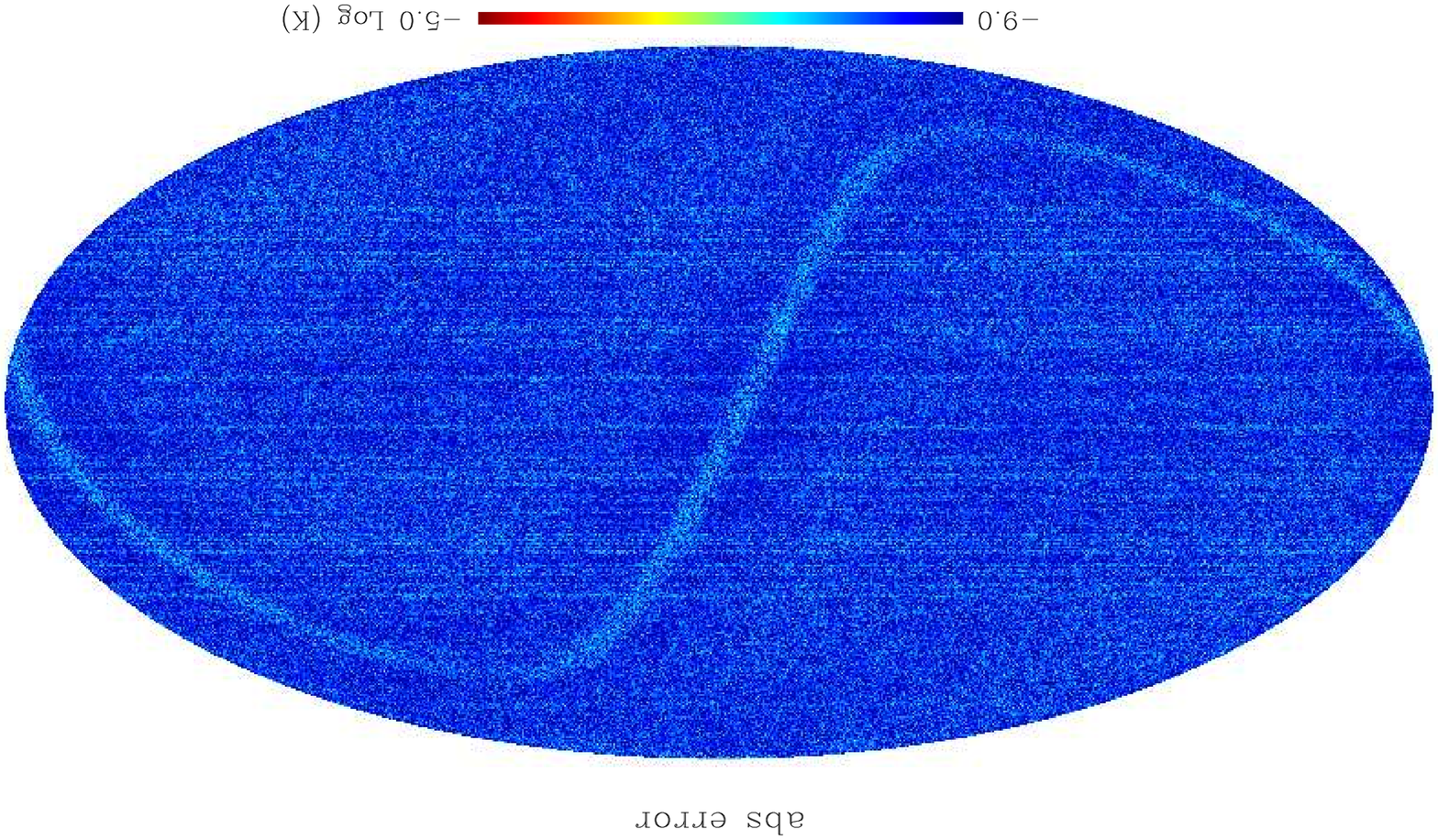}\\
\end{tabular}
\caption{\label{fig:map_making_error}\textit{Top panel}: The input CMB
  realization directly convolved with the axisymmetric component of the
  elliptical Gaussian beam used for the satellite-like
  simulation. \textit{Second panel}: The absolute residuals between the
  isotropically smoothed map shown in the top panel and the simple
  binned map constructed from the simulated
  TOD. \textit{Lower two panels}: The absolute residuals between the
  isotropically smoothed map shown in the top panel and maps made
  using the algorithm described in Section~\ref{sec:map_making} with
  $k_{\rm  max}{=}2$ and $k_{\rm  max}{=}4$. The residuals in the $k_{\rm max} {=}4$ case
  (\emph{bottom panel}) are comparable with the numerical noise of our
  convolution code.}
\end{center}
\end{figure*}

To estimate the temperature map, the algorithm of
Section~\ref{sec:map_making} extracts only the spin-0 components of the
TOD. This quantity contains the temperature of the CMB smoothed with
the axisymmetric component of the beam. To demonstrate this we
simulate a noise free TOD using the elliptical Gaussian and the
satellite-like scan strategy. We plot the CMB used in the simulation
smoothed with the axisymmetric component of the beam in the top panel
of Fig.~\ref{fig:map_making_error}. The other three panels in this
figure show the absolute error between this isotropically smoothed map
and a simple binned map (second panel); and between the isotropically
smoothed map and maps produced using the map-making algorithm with $k_{\rm
  max}{=}2$ and $k_{\rm max}{=}4$ (lower two panels). The reduction in
the bias due to beam asymmetry as $k_{\rm max}$ is increased is
clearly demonstrated by this figure.

We also tested the ability of the map-making algorithm to remove the
leakage from temperature to polarisation due to the asymmetry of the
beam. We performed this test using the approach described in
Section~\ref{sec:TtoP_rm} which makes use of a previously estimated
temperature map obtained with the method described in
Section~\ref{sec:ext_spin_pix} with $k_{\rm max}{=}4$. In our
simulation we have not included an input polarisation signal so the
expected power in the polarisation maps will be zero in the absence of
systematics. As described in Section~\ref{sec:TtoPleak}, the
temperature to polarization leakage could contaminate either the
$E$-mode or the $B$-mode power spectrum depending on the relative
orientation between the beam ellipticity and the polarisation response
of the detector.

\begin{figure}
\begin{center}

\includegraphics[width=0.45\textwidth,angle=180]{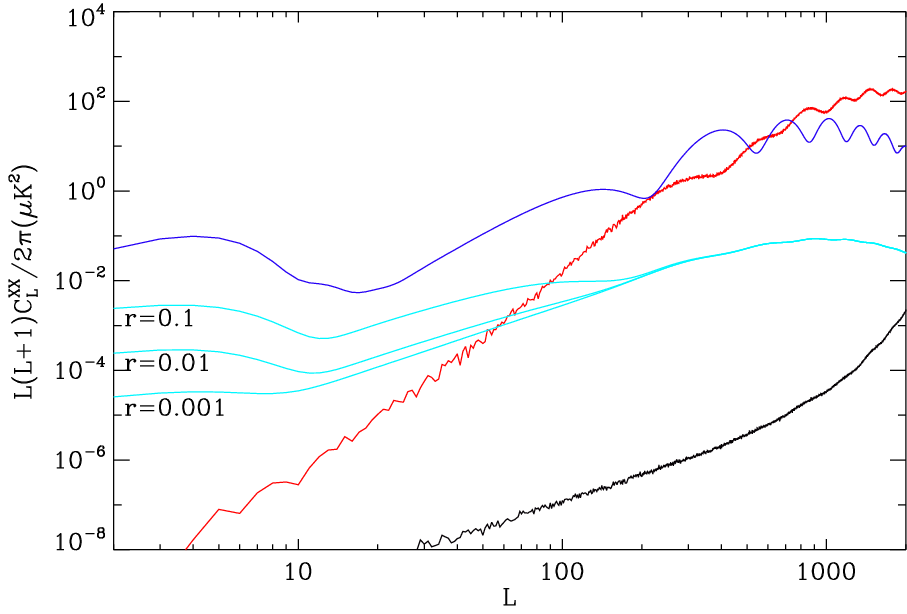}

\caption{The power spectrum of the temperature signal leaked to
  polarization as a result of beam asymmetries for the simulated
  satallite-like experiment (red). This power could manifest itself as
  either an $E$-mode, a $B$-mode or a combination of both -- see the
  discussion in Section~\ref{sec:TtoPleak}. The black line shows the
  power spectrum of the polarisation leakage after applying the
  correction algorithm described in Section~\ref{sec:TtoP_rm}. For
  comparison we plot the expected $E$-mode and $B$-mode signal in dark
  and light blue respectively. The expected $B$-mode signal is plotted
  for three cases of the tensor-to-scalar ratio:
  $r{=}0.001,0.01,0.1.$}\label{fig:cl_pol_clean}
\end{center}
\end{figure}

In Fig.~\ref{fig:cl_pol_clean} we plot the power of the leaked
temperature to polarisation signal and the power in the polarization
maps after this leakage has been removed using the appproach described in
Section~\ref{sec:TtoP_rm}. The power in the cleaned polarisation map
has been successfully reduced by $\sim$4 orders of magnitude compared
to the uncleaned map.

\section{Discussion} \label{sec:discuss}

We have developed two separate techniques to remove asymmetry bias in
CMB experiments. The first is a pseudo-$C_{\ell}$ method that corrects
for the asymmetry bias on the temperature and polarisation power
spectra. There is no assumption on the scan strategy and only a modest
approximation to the beam is required. The analysis is similar to the
MASTER algorithm \citep{2002ApJ...567....2H} and its extension to
polarisation \citep{2005MNRAS.360.1262B}. However we include
additional contributions to the coupling operator that corrects for
the contaminating effects of the beam asymmetry. We note that this
algorithm can only be applied to experiments that can directly measure
the $Q$ and $U$ Stokes parameters in the timeline, e.g. a differencing
experiment with both instrument-$Q$ and instrument-$U$ detectors on
its focal plane. This is due to the fact that we make no attempt to
make a map and so we must have measurements of the Stokes parameters
in the telescope's frame of reference. At present the formalism
assumes the four beams required for such a differencing experiment are
the same. In future work we will relax this assumption.

Using this formalism we also examined the inter-spectra coupling
resulting from the beam asymmetry. As an example, we investigated the
coupling between temperature and B-mode power in the presence of an
ideal scan strategy. In doing so we showed that the temperature power
will be coupled to the B-mode power if the beam asymmetry is at an
angle to the polarisation sensitivity direction of the detector. This
result is in agreement with previous work \citep{2008PhRvD..77h3003S}.

The coupling operator in equation \eqref{eq:oper} allows us to
calculate the contribution to each pseudo-$C_{\ell}$ from the sky
power, due to both the mask and beam asymmetry. By inverting this
operator we can, therefore, calculate an unbiased estimate for the CMB
power spectra. To calculate the operator $\vect{O}_{i_1 i_2}$ we must
cap the azimuthal dependence of the beam, $k_{\rm max}{\lesssim}
10$. We demonstrated in Section~\ref{sec:reducesum}, and in
Figs.~\ref{fig:beam1} and \ref{fig:gaus_decomp}, that
a simulated beam from a multi-moded horn, and an elliptical Gaussian,
equation~\eqref{eq:gaus_beam}, are both well described by only a few azimuthal
modes. This property allows us to remove to a high degree of accuracy
the bias due to the beam asymmetry for any given scan strategy. 
%This is because the coupling operator has the scan strategy encoded
%into it.

We then went on to implement this algorithm in the temperature only
case for two simulated experimental set-ups: a balloon-like experiment and a
satellite-like experiment. We showed that we could successfully
recover the input power spectrum, when the TOD was created using a
highly asymmetric beam in the presence of both a severe sky-cut
and instrument noise. These tests showed that the algorithm
can deal comfortably with beams that have much higher levels of
asymmetry than that typically found in real CMB experiments, and using
two completely different and realistic scan strategies. The scan
strategies investigated include the proposed scanning mode for the
forthcoming balloon-borne LSPE experiment~\citep{2012arXiv1208.0281T},
and a scan strategy optimized for a possible future CMB polarization
satellite mission~\citep{2009arXiv0906.1188B}.

The second technique that we propose for removing beam asymmetry bias
is a new map-making algorithm. The map-making algorithm produces
temperature and polarisation maps of the CMB that are smoothed with
only the axisymmetric component of the beam. It achieves this by
separating out the different spin components of the detected signal
using equation \eqref{eq:spin_sep}. This allows us to obtain unbiased
estimates for the spin-0 and spin-2 components of the signal which
correspond to the temperature and polarisation of the pixel. The
temperature map estimated using this technique will be clean of
systematics due to the asymmetry of the beam. However, the
polarisation map will still contain a temperature to polarisation
leakage term due to the $k{=}2$ azimuthal mode of the beam. This
contaminating signal can effectively be removed by calculating the
leakage from the estimated temperature map and the known beam response
as described in Section~\ref{sec:TtoP_rm}.

Removing the effects of beam asymmetry at the map level is preferable
to removing it using the pseudo-$C_{\ell}$ method. There are three
main reasons for this. Firstly there will be no aliasing between
spectra. By removing the inter-spectra leakage at the power spectrum
level we will increase the statistical error on the recovered power
spectra by a factor proportional to the amplitude of the leaked
spectra. A more optimal estimate of the true power spectrum can
therefore be obtained if we can remove the leakage in the map
domain. Secondly, the map-making algorithm can be applied before
foregrounds are removed, allowing the application of traditional
foreground removal techniques on the beam asymmetry-cleaned
maps. Thirdly, the cleaned maps can subsequently be used for other
science applications beyond the power spectrum such as CMB lensing,
non-Gaussianity studies and foreground science. {\color{black}Note
error propagation from uncertainty in the beam shape may be more
complicated. We propose that an experiment should Monte-Carlo
over uncertainties in the beam to be able to fully understand the
errors on the map and therefore the science.}

The map-making algorithm requires a scan strategy to cross a pixel at
multiple orientations. If this is not the case, the matrix $H_{kk'}$,
defined in equation \eqref{eq:Hbig}, is not invertible. In such cases,
the map-making method cannot be applied (although the pseudo-C$_\ell$
approach could still be used). We note that, as an alternative to a
highly redundant scan strategy, the required polarisation angle
coverage could also be provided through the use of an appropriately
positioned HWP, as is the case in the LSPE
experiment~\citep{2012arXiv1208.0281T}.

%This is the case for the scan
%strategy we investigated for the simulated balloon-like
%experiment. Fig.~\ref{fig:extreme_tod} shows the $p_2$ quality,
%\eqref{eq:p_qual} of this scan pattern which is poor and does not
%provide enough crossing angles to apply to map-making method. 
%However the pseudo-$C_{\ell}$ approach would be suitable for such an
%experiment. It should be mentioned that in balloon experiments a
%rotating HWP can increase the crossing angles and mitigate asymmetry
%bias, this is not something we have not included in our simulations.

In Section~\ref{sec:sim_test_map} we showed that we could use the map-making
 algorithm to correctly recover an input temperature power
spectrum free of asymmetry bias for the same simulated satellite-like
experiment as was used to test the pseudo-$C_{\ell}$ technique. In
doing this we found that the noise level in the recovered power
spectrum increased by ${\sim}20\%$ compared to that recovered from a
simply binned map. We also demonstrated that we could successfully
correct the temperature and polarisation signal for the effects of
beam asymmetry at the map level. To demonstrate this, we simulated a
satellite-like experiment free of noise and we computed the
temperature and polarisation maps using the algorithm described in
Section~\ref{sec:map_making}. For the case of $k_{\rm max}{=}4$ we
showed that the residual error between the resulting estimated
temperature map and the input temperature map smoothed with the
axisymmetric component of the beam was comparable to the numerical
error in the convolution code. Finally, we have also demonstrated that
this method could be very powerful for correcting for temperature to
polarization leakage due to beam asymmetries. This ability,
which is demonstrated in Fig.~\ref{fig:cl_pol_clean}, could prove
extremely useful for controlling systematics due to beam asymmetries
in future precision CMB $B$-mode polarization experiments.

\section*{Acknowledgments}

CGRW acknowledges the award of a STFC quota studentship. MLB is
grateful to the European Research Council for support through the
award of an ERC Starting Independent Researcher Grant (EC FP7 grant
number 280127). MLB also thanks the STFC for the award of Advanced and
Halliday fellowships (grant number ST/I005129/1). We also thank
 Bruno Maffei and the LSPE collaboration for helpful
discussions. Some of the results in this paper have been derived using
the {\sevensize HEALPix}~\citep{2005ApJ...622..759G} package.
 
\bibliographystyle{mn2e}
\bibliography{ref}

\onecolumn

\appendix

\section{Relationship between our pseudo-$C_{\ell}$ and the standard pseudo-$C_{\ell}$} \label{sec:rel_to_stand}

Here we show that the pseudo-$C_{\ell}$ expression derived in
Section~\ref{sec:pseudo-cl-technique} reduces to a rewriting of the
standard polarised pseudo-$C_{\ell}$ presented in
\citet{2005MNRAS.360.1262B}, in the case where the beam is
axisymmetric and where the experiment has both instrument-$Q$ and
instrument-$U$ detectors. We begin by looking at
the coupling kernel,

\begin{align}
K_{m_1 k_1 m_2 k_2}^{\ell_1 \ell_2} &\equiv \int d^3\bom D^{\ell_1}_{m_1 k_1}(\bomega) D^{\ell_2 *}_{m_2 k_2}(\bomega) W(\bomega)n(\bom).\\
\end{align}
{\color{black} 	We use the weighting function $n(\bom)$ to apodise the hit map and
apply a galactic mask. While, in general, it can be a function of
$\theta$, $\phi$ and $\psi$, in practice, it is sufficient for it to
be a function of just $\theta$ and $\phi$. The function will be discretised.
 It should be chosen such that $\int d\psi W(\bom)n(\theta,\phi)$ for a 
given pixel ranges from 0 to 1 and acts as
the standard window function in the pseudo-$C_\ell$. }
The requirement that the experiment has both an instrument-$Q$ and
-$U$ detector means that for every orientation of the telescope $\bom$
we have 4 detections of the sky each at $[\theta, \phi, \psi +
  j\pi/4]$, where j{=}0,1,2,3. A consequence of this arrangement of
detectors is that

\begin{align}
\int d\psi W(\bom)e^{ik\psi} = 0 \:\:\:\text{for}\:\:\: k = \pm2,\pm4.
\end{align}
Therefore $ K_{m_1 k_1 m_2 k_2}^{\ell_1 \ell_2}{=}0$ if
$k_1-k_2{=}\pm2,\pm4$. We can now examine the Wigner decomposition of
the time stream, in particular the $k{=}0,\pm2$ components, as these will contain
information on the temperature and polarisation of the CMB. We also
assume the beam to be axisymmetric and have a zero cross polar
response, so that $b_{s\ell k}{=}0$ for $s{\ne}{-}k$. We make
this assumption in order to make the connection with the standard
pseudo-$C_{\ell}$ approach. Applying these constraints to equation
\eqref{eq:TfromK} we get,

\ba
T^{\ell_1*}_{m_1 0} &=& \sum_{\ell_2 m_2} b^*_{0 \ell_2 0}a_{0 \ell_2 m_2} K_{m_1 0 m_2 0}^{\ell_1 \ell_2}\\
T^{\ell_1*}_{m_1 \pm2} &=& \sum_{\ell_2 m_2}b^*_{\mp2 \ell_2 \pm2}a_{\mp2 \ell_2 m_2} K_{m_1 \pm2 m_2 \pm2}^{\ell_1 \ell_2}.
\ea
Using these expressions, we can show that $E$- and $B$-mode-like
decompositions (equations~\ref{eq:TlmE} \& \ref{eq:TlmB}) of the TOD in this experiment will be,

\ba
T^{\ell_1*}_{m_1 E} &=& \sum_{\ell_2 m_2}b^*_{-2 \ell_2
  2}\left(a^E_{\ell_2 m_2} K_{\ell_1 \ell_2 m_1 m_2}^{+} + a^B_{\ell_2
  m_2} K_{\ell_1 \ell_2 m_1 m_2}^{-}\right)\nonumber \\
T^{\ell_1*}_{m_1 B} &=& \sum_{\ell_2 m_2}b^*_{-2 \ell_2 2}\left(a^B_{\ell_2 m_2} K_{\ell_1 \ell_2 m_1 m_2}^{+} - a^E_{\ell_2 m_2} K_{\ell_1 \ell_2 m_1 m_2}^{-}\right),
\label{eq:app_e_b_like}
\ea
where we have defined,

\ba
K_{\ell_1 \ell_2 m_1 m_2}^{+} &\equiv& -\frac{1}{2}\left(K_{m_1 2 m_2 2}^{\ell_1 \ell_2} + K_{m_1 -2 m_2 -2}^{\ell_1 \ell_2}\right)\\
K_{\ell_1 \ell_2 m_1 m_2}^{-} &\equiv& \frac{i}{2}\left(K_{m_1 2 m_2 2}^{\ell_1 \ell_2} - K_{m_1 -2 m_2 -2}^{\ell_1 \ell_2}\right).
\ea
Comparing the expressions of equation~\eqref{eq:app_e_b_like} with
equation~(10) of \citet{2005MNRAS.360.1262B} shows that the $E$- and
$B$-mode-like decompositions of the TOD defined in equations
\eqref{eq:TlmE} \& \eqref{eq:TlmB} have a similar form to the $E$- and
$B$-mode decompositions of a polarisation map used in the standard
pseudo-$C_{\ell}$ technique. They are similar in the sense that they
are both the sky polarisation smoothed with the beam, and then
convolved with a window function. They differ only in normalisation
factors. As the decompositions are similar, the pseudo-$C_{\ell}$
constructed from them will also be similar.

\section{Product of two coupling kernels} \label{sec:productkern}

We require the product of two coupling kernels. In order to calculate this, we start from the definition the Kernel

\begin{align}
K_{m_1 k_1 m_2 k_2}^{\ell_1 \ell_2} &\equiv \int d^3\bom D^{\ell_1}_{m_1 k_1}(\bomega) D^{\ell_2 *}_{m_2 k_2}(\bomega) W(\bomega)n(\bom)\\
&=\sum_{\ell_3 m_3 k_3}w^{\ell_3}_{m_3 k_3} \int d^3\bom D^{\ell_1}_{m_1 k_1}(\bomega) D^{\ell_2 *}_{m_2 k_2}(\bomega) D^{\ell_3}_{m_3 k_3}(\bomega)\\
&=8 \pi^2 (-1)^{m_2 + k_2}\sum_{\ell_3 m_3 k_3}w^{\ell_3}_{m_3 k_3}  \threej{\ell_1}{\ell_2}{\ell_3}{m_1}{-m_2}{m_3}\threej{\ell_1}{\ell_2}{\ell_3}{k_1}{-k_2}{k_3},
\end{align}
where the second equality comes from an identity found in \citet{varshalovich:etal:1988}. We can now evaluate the product summed over certain indices $m_1$, and $m_2$

\begin{align}
M^{\ell_1 \ell_2}_{k_1 k_1' k_2 k_3} &\equiv \sum_{m_1 m_2} K_{m_1 k_1 m_2 k_2}^{\ell_1 \ell_2} K_{m_1 k_1' m_2 k_3}^{\ell_1 \ell_2*} \\
&= 64\pi^4 (-1)^{k_2 +k_3}\!\!\!\!\sum_{\substack{m_1 m_2 {}\\ \ell_4 m_4 k_4 {}\\ \ell_5 m_5 k_5}}\!\!\!w^{\ell_4}_{m_4 k_4}w^{\ell_5 *}_{m_5 k_5} \threej{\ell_1}{\ell_2}{\ell_4}{m_1}{-m_2}{m_4}\threej{\ell_1}{\ell_2}{\ell_4}{k_1}{-k_2}{k_4} \nonumber\\
&\:\:\:\:\:\:\:\:\:\:\:\:\:\:\:\:\:\:\:\:\:\:\:\:\:\:\:\:\:\:\:\:\:\:\:\:\:\:\:\:\:\:\:\:\:\:\:\:\:\:\:\:\:\:\:\:\:\:\:\:\:\:\:\:\:\:\:\:\:\:\:\:\:\:\:\:\:\:\:\:\:\:\:\:\times \threej{\ell_1}{\ell_2}{\ell_5}{m_1}{-m_2}{m_5}\threej{\ell_1}{\ell_2}{\ell_5}{k_1'}{-k_3}{k_5}.
\end{align}
The Wigner 3$j$ orthogonality relation is 

\begin{align}
\sum_{m_1 m_2}\threej{\ell_1}{\ell_2}{\ell_4}{m_1}{m_2}{m_4}\threej{\ell_1}{\ell_2}{\ell_5}{m_1}{m_2}{m_5} = \frac{1}{2\ell_4 +1} \delta_{\ell_4, \ell_5} \delta_{m_4, m_5},
\end{align}
which enables the sum over $m_1$ and $m_2$ to be performed and evaluating the Kronecker $\delta$ gives
\begin{align}
M^{\ell_1 \ell_2}_{k_1 k_1' k_2 k_3} &= 64\pi^4 (-1)^{k_2 +k_3}\!\!\!\!\!\! \sum_{\ell_4 m_4 k_4 k_5} \!\!\!\!\!\!w^{\ell_4}_{m_4 k_4}w^{\ell_4 *}_{m_4 k_5} \frac{1}{2\ell_4 +1} \threej{\ell_1}{\ell_2}{\ell_4}{k_1}{-k_2}{k_4} \threej{\ell_1}{\ell_2}{\ell_4}{k_1'}{-k_3}{k_5}.
\end{align}
We can simplify this further by defining the window correlation matrix $\mathcal{W}^\ell_{k_1, k_2} \equiv \sum_{m} w^{\ell}_{m k_1}w^{\ell *}_{m k_2}$. Also we can use the selection rule in the 3$j$ symbols that states that the sum of the bottom row must be equal to zero for the symbol to be non-zero. This gives
us

\begin{align}
M^{\ell_1 \ell_2}_{k_1 k_1' k_2 k_3} &= 64\pi^4 (-1)^{k_2 +k_3}\!\! \sum_{\ell_4 k_4 k_5} \!\! \mathcal{W}^{\ell_4}_{k_4, k_5} \frac{1}{2\ell_4 +1} \threej{\ell_1}{\ell_2}{\ell_4}{k_1}{-k_2}{k_4} \threej{\ell_1}{\ell_2}{\ell_4}{k_1'}{-k_3}{k_5}\delta_{k_4,k_2-k_1}\delta_{k_5,k_3-k_1'} \nonumber \\
&= 64\pi^4 (-1)^{k_2 +k_3}\sum_{l_4}  \mathcal{W}^{\ell_4}_{k_2-k_1, k_3-k_1'} \frac{1}{2\ell_4 +1} \threej{\ell_1}{\ell_2}{\ell_4}{k_1}{-k_2}{k_2-k_1} \threej{\ell_1}{\ell_2}{\ell_4}{k_1'}{-k_3}{k_3-k_1'}. \label{eq:calM}
\end{align}

\section{Symmetries in $M^{\lowercase{\ell_1 \ell_2}}_{\lowercase{k_1 k_1' k_2 k_3}}$} \label{sec:symM}

The symmetries in the matrix $M^{\ell_1 \ell_2}_{k_1 k_1' k_2 k_3}$ are important to reduce the computation time. First we look at swapping the indices $k_2$ and $k_3$ along with $k_1$ and $k_1'$:

\begin{align}
M^{\ell_1 \ell_2}_{k_1' k_1 k_3 k_2} &= 64\pi^4 (-1)^{k_2 +k_3}\sum_{\ell_4}  \mathcal{W}^{\ell_4}_{k_3-k_1', k_2-k_1} \frac{1}{2\ell_4 +1} \threej{\ell_1}{\ell_2}{\ell_4}{k_1'}{-k_3}{k_3-k_1'} \threej{\ell_1}{\ell_2}{\ell_4}{k_1}{-k_2}{k_2-k_1} \nonumber \\
&=64\pi^4 (-1)^{k_2 +k_3}\sum_{\ell_4}  \mathcal{W}^{\ell_4*}_{k_2-k_1, k_3-k_1'} \frac{1}{2\ell_4 +1} \threej{l_1}{l_2}{l_4}{k_1}{-k_2}{k_2-k_1} \threej{\ell_1}{\ell_2}{\ell_4}{k_1}{-k_3}{k_3-k_1'}\nonumber \\
&=M^{\ell_1 \ell_2*}_{k_1 k_1' k_2 k_3},
\end{align}
where we have used the definition of window correlation matrix to say that $\mathcal{W}^\ell_{k_2, k_1}=\mathcal{W}^{\ell*}_{k_1, k_2}$.
Now we look at the reversing the sign of all $k_i$. This implies that

\begin{align}
\mathcal{W}^\ell_{-k_1, -k_2} &= \sum_{m} w^{\ell}_{m -k_1}w^{\ell *}_{m -k_2}\nonumber \\
&= \sum_{m} (-1)^{m+k_1}w^{\ell*}_{-m k_1}(-1)^{m+k_1}w^{\ell}_{-m k_2}\nonumber \\
&= (-1)^{k_1 + k_2}\mathcal{W}^{\ell*}_{k_1, k_2}.
\end{align}

Now we can show symmetry in the matrix $M$ is given by

\begin{align}
M^{\ell_1 \ell_2}_{-k_1 -k_1' -k_2 -k_3} &= 64\pi^2 (-1)^{k_2 +k_3}\sum_{\ell_4}  \mathcal{W}^{\ell_4}_{-k_2+k_1, -k_3+k_1'} \frac{1}{2\ell_4 +1} \threej{\ell_1}{\ell_2}{\ell_4}{-k_1}{+k_2}{-k_2+k_1} \threej{\ell_1}{\ell_2}{\ell_4}{-k_1'}{+k_3}{-k_3+k_1'}\nonumber \\
 &= 64\pi^4 (-1)^{k_2 +k_3}\sum_{\ell_4}  (-1)^{k_2 -k_1+k_3-k_1' }\mathcal{W}^{\ell_4*}_{k_2-k_1, k_3-k_1'} \frac{1}{2\ell_4 +1}\threej{\ell_1}{\ell_2}{\ell_4}{k_1}{-k_2}{k_2-k_1}\threej{\ell_1}{\ell_2}{\ell_4}{k_1'}{-k_3}{k_3-k_1'}\nonumber \\
 &= (-1)^{k_2 -k_1+k_3-k_1'}M^{\ell_1 \ell_2*}_{k_1 k_1' k_2 k_3},
\end{align}
where in the second equality we have used the relation

\begin{align}
\threej{\ell_1}{\ell_2}{\ell_3}{m_1}{m_2}{m_3}\threej{\ell_1}{\ell_2}{\ell_3}{m_1'}{m_2'}{m_3'}=\threej{\ell_1}{\ell_2}{\ell_3}{-m_1}{-m_2}{-m_3}\threej{\ell_1}{\ell_2}{\ell_3}{-m_1'}{-m_2'}{-m_3'}.
\end{align}

\section{Explicit form of the coupling operator}\label{app:coup_op}

In section \ref{sec:est_cmb} we use the coupling operator $\vect{O}_{i_1 i_2}$. Here we explicitly write it in terms of the sub operators $O_{\ell_1 \ell2}^{k_1 k_1' s_1 s_2}$

\ba
\vect{O}_{ij} = {\setstretch{1.5}\left( \begin{array}{cccccc}
O_{\ell_1 \ell_2}^{0000} & O_{\ell_1 \ell_2}^{0002} & O_{\ell_1 \ell_2}^{000-2} & O_{\ell_1 \ell_2}^{0022} & O_{\ell_1 \ell_2}^{002-2} &O_{\ell_1 \ell_2}^{00-2-2}\\
O_{\ell_1 \ell_2}^{0200} & O_{\ell_1 \ell_2}^{0202} & O_{\ell_1 \ell_2}^{020-2} & O_{\ell_1 \ell_2}^{0222} & O_{\ell_1 \ell_2}^{022-2} &O_{\ell_1 \ell_2}^{00-2-2}\\
O_{\ell_1 \ell_2}^{0-200} & O_{\ell_1 \ell_2}^{0-202} & O_{\ell_1 \ell_2}^{0-20-2} & O_{\ell_1 \ell_2}^{0-222} & O_{\ell_1 \ell_2}^{0-22-2} &O_{\ell_1 \ell_2}^{0-2-2-2}\\
O_{\ell_1 \ell_2}^{2200} & O_{\ell_1 \ell_2}^{2202} & O_{\ell_1 \ell_2}^{220-2} & O_{\ell_1 \ell_2}^{2222} & O_{\ell_1 \ell_2}^{222-2} &O_{\ell_1 \ell_2}^{22-2-2}\\
O_{\ell_1 \ell_2}^{2-200} & O_{\ell_1 \ell_2}^{2-202} & O_{\ell_1 \ell_2}^{2-20-2} & O_{\ell_1 \ell_2}^{2-222} & O_{\ell_1 \ell_2}^{2-22-2} &O_{\ell_1 \ell_2}^{2-2-2-2}\\
O_{\ell_1 \ell_2}^{-2-200} & O_{\ell_1 \ell_2}^{-2-202} & O_{\ell_1 \ell_2}^{-2-20-2} & O_{\ell_1 \ell_2}^{-2-222} & O_{\ell_1 \ell_2}^{-2-22-2} &O_{\ell_1 \ell_2}^{-2-2-2-2}\end{array} \right)}.\nonumber
\ea

\section{Relations between $C^{ss'}_{\ell}$ and $C^{XY}_{\ell}$} \label{app:spin2XY}

The following matrix operation allows conversion from $C^{ss'}_{\ell}$ to $C^{XY}_{\ell}$,

\ba
\left( \begin{array}{c}
C_{\ell}^{TT}\\
C_{\ell}^{TE}\\
C_{\ell}^{TB}\\
C_{\ell}^{EE}\\
C_{\ell}^{EB}\\
C_{\ell}^{BB}\end{array}\right)
= \left(\begin{array}{ccccccc}
1 & 0 & 0 & 0 & 0 & 0  & 0 \\
0 & -1/2 & -1/2 & 0 & 0 & 0  & 0 \\
0 & i/2 & -i/2 & 0 & 0 & 0  & 0 \\
0 & 0 & 0 & 1/4 & 1/4 & 1/4  & 1/4 \\
0 & 0 & 0 & -i/4 & i/4 & -i/4  & i/4 \\
0 & 0 & 0 & -1/4 & -1/4 & 1/4  & 1/4  \end{array}\right)
\left( \begin{array}{c}
C_{\ell}^{00}\\
C_{\ell}^{02}\\
C_{\ell}^{0-2}\\
C_{\ell}^{22}\\
C_{\ell}^{2-2}\\
C_{\ell}^{-22}\\
C_{\ell}^{-2-2}\end{array}\right).
\ea
The matrix is not square, this does not imply that there is more information in the right than the left. There are the same number of degrees of freedom on both sides this is because $C_{\ell}^{2-2}{=}C_{\ell}^{-22}$.

%\section{Sky coverage for simulation tests} \label{sec:cover}
%
%Here we show the sky coverage for the tests described in \S\ref{sec:sim_test}
%
%\input{sections/sky_cover_fig}

\label{lastpage}

\end{document}